\documentclass[10pt,a4paper,english,pre,onecolumn,superscriptaddress,floatfix]{revtex4-1}
\usepackage[T1]{fontenc}
\usepackage[latin9]{inputenc}
\usepackage[a4paper]{geometry}
\usepackage{amsmath}
\usepackage{amssymb}
\usepackage{graphicx}

\usepackage{color}

\makeatletter
\@ifundefined{textcolor}{}
{%
 \definecolor{BLACK}{gray}{0}
 \definecolor{WHITE}{gray}{1}
 \definecolor{RED}{rgb}{1,0,0}
 \definecolor{GREEN}{rgb}{0,1,0}
 \definecolor{BLUE}{rgb}{0,0,1}
 \definecolor{CYAN}{cmyk}{1,0,0,0}
 \definecolor{MAGENTA}{cmyk}{0,1,0,0}
 \definecolor{YELLOW}{cmyk}{0,0,1,0}
}

\makeatother

\usepackage{babel}
\begin{document}

\title{The zero-patient problem with noisy observations}

\author{Fabrizio~Altarelli}
\affiliation{DISAT and Center for Computational Sciences, Politecnico di Torino,
Corso Duca degli Abruzzi 24, 10129 Torino, Italy}
\affiliation{Collegio Carlo Alberto, Via Real Collegio 30, 10024 Moncalieri, Italy}

\author{Alfredo~Braunstein}
\affiliation{DISAT and Center for Computational Sciences, Politecnico di Torino,
Corso Duca degli Abruzzi 24, 10129 Torino, Italy}
\affiliation{Human Genetics Foundation, Via Nizza 52, 10126 Torino, Italy }
\affiliation{Collegio Carlo Alberto, Via Real Collegio 30, 10024 Moncalieri, Italy}

\author{Luca~Dall'Asta}
\affiliation{DISAT and Center for Computational Sciences, Politecnico di Torino,
Corso Duca degli Abruzzi 24, 10129 Torino, Italy}
\affiliation{Collegio Carlo Alberto, Via Real Collegio 30, 10024 Moncalieri, Italy}

\author{Alessandro~Ingrosso}
\affiliation{DISAT and Center for Computational Sciences, Politecnico di Torino,
Corso Duca degli Abruzzi 24, 10129 Torino, Italy}

\author{Riccardo~Zecchina}
\affiliation{DISAT and Center for Computational Sciences, Politecnico di Torino,
Corso Duca degli Abruzzi 24, 10129 Torino, Italy}
\affiliation{Human Genetics Foundation, Via Nizza 52, 10126 Torino, Italy }
\affiliation{Collegio Carlo Alberto, Via Real Collegio 30, 10024 Moncalieri, Italy}

\begin{abstract}
A Belief Propagation approach has been recently proposed for
the zero-patient problem in a SIR epidemics. The zero-patient problem consists in finding the 
initial source of an epidemic outbreak given observations at a later time. 
In this work, we study a harder but related
inference problem, in which observations are noisy and
there is confusion between observed states. In addition to studying the zero-patient 
problem, we also tackle the problem of completing and correcting the observations 
possibly finding undiscovered infected individuals and false test results.

Moreover, we devise a set of equations, based on the variational expression of the 
Bethe free energy, to find the zero patient along with maximum-likelihood
epidemic parameters. We show, by means of simulated epidemics, how this method 
is able to infer details on the past history of an epidemic outbreak based solely on the 
topology of the contact network and a single snapshot of partial and noisy observations.
\end{abstract}
\maketitle

\section{Introduction}

\global\long\def\dd{\partial}
\global\long\def\I{\mathbb{I}}
\global\long\def\sign{\mbox{sign}}
 Epidemic compartment models provide a simple and useful mathematical
description of the mechanisms behind disease transmission between
individuals, in which only the most prominent ingredients are included
\cite{bailey_mathematical_1975}. One of the most celebrated of these models,
Susceptible-Infected-Recovered (SIR)\cite{kermack1927contribution},
describes diseases in which the person contracting the disease becomes
immune to future infections after recovery, such as measles, rubella,
chicken pox and generic influenza. The same model can be applied to
lethal diseases, such as HIV or Ebola, provided that the recovered
state is replaced by a removed state. Being simple and mathematically
appealing, models such as SIR can be employed to study under which
conditions and with which consequences large epidemic outbreaks can
occur. For most examples of infective diseases, contagion runs over
a network of effective contacts between individuals. These contact
networks have been inaccessible for decades, but thanks to recent
advances in technology miniaturization (e.g. by means of RFID-endowed
badges to signal the proximity between individuals) and the popularization
of the Internet (e.g. for the construction of databases of self-reported
interactions), at least in simple and controlled scenarios the interaction
patterns of individual contacts can be almost entirely reconstructed
\cite{isella_whats_2011,rocha_information_2010}. Modern computational
epidemiology can thus rely on accurate data and on powerful computers
to run large-scale simulations of stochastic compartment models on
real contact networks \cite{danon2011networks,marathe_2013}. 

In addition to epidemic forecast and control, a problem that has gained
attention in recent years is the one of reconstructing the history
of an epidemic outbreak \cite{shah_detecting_2010,comin_identifying_2011,shah_rumors_2011,fioriti2012predicting,pinto_locating_2012,antulov-fantulin_statistical_2013,dong2013rooting,lokhov_inferring_2013,luo_identifying_2013,zhu_information_2013,franceschetti2013,altarelli_prl_2014},  as e.g. the path of contagion to a specific infected individual.
In particular, identifying the origin (or the set of {\em seeds} or {\em sources}) of an epidemic outbreak
in the general case is an open problem, even assuming
simple discrete-time stochastic epidemic models, such as the SI and
SIR model. The reason becomes clear when the inference is formulated
as a maximum likelihood estimation problem. Estimating the maximum
of a properly defined likelihood function corresponds to solve a (generally
non-convex) optimization problem in the space of all possible epidemic
propagations that are compatible with the data. For propagations with
a unique source on regular trees, a maximum likelihood estimator was
proposed by Shah and Zaman \cite{shah_detecting_2010,shah_rumors_2011}
under the name of rumor centrality (see also \cite{dong2013rooting})
and extended to probabilistic observations in Ref.~\cite{franceschetti2013}.
For a specific continuous-time epidemic process, an optimal estimator
on general trees was put forward by Pinto et al. \cite{pinto_locating_2012}.
On general graphs, the number of propagation paths grows exponentially
with the number of nodes making the exact inference unfeasible in
practice. Instead of evaluating the likelihood function, Zhu and Ying
put forward a method to select the path that most likely leads to
the observed snapshot \cite{zhu_information_2013}. For general graphs,
other heuristic inference methods are based on centrality measures
\cite{comin_identifying_2011,fioriti2012predicting}, on the distance
between observed data and typical outcome of propagations for given
initial conditions \cite{antulov-fantulin_statistical_2013} or on
the assumption that the epidemic propagation follows a breadth-first
search tree \cite{pinto_locating_2012,luo_identifying_2013}.
Even fewer results exist for epidemic inference with multiple sources
\cite{luo_identifying_2013}. A message-passing approach for the computation of epidemic dynamics
was first introduced by Karrer and Newman \cite{karrer_message_2010}
and applied to source detection in Ref. \cite{lokhov_inferring_2013}, with 
a further mean-field approximation of the likelihood function.

Recently, a Belief Propagation (BP) approach was proposed for the Bayesian
inference of the origin of an epidemics \cite{altarelli_prl_2014}. The main
idea is that of exploiting a graphical model representation of the stochastic
dynamics of the SI and SIR models to devise an efficient message-passing
algorithm for the evaluation of the posterior distribution of the epidemic
sources.  The BP approach is exact on trees and it works very well also on
general graphs, outperforming other methods on many graph topologies, in the
presence of one or more sources, and also when observations are by large extent
incomplete. In this work, we build on the BP approach by studying a harder
variant of the inference problem in which either (a) it is not possible to
distinguish between recovered or susceptible individuals or (b) the observation
is noisy, i.e. there is a non-zero probability of making an error
in the observation of each individual. The first case was already
described in Ref. \cite{zhu_information_2013} using the most likely
infection path method on trees and similar heuristics on general graphs.
The second case was not directly faced in the literature, although
the problem of determining the causative network of epidemiological
data in the presence of false negatives and positives has recently
attracted some attention \cite{milling2013detecting,meirom2014localized}.
We will show that the BP approach allows for more complete inference, as e.g. 
(a) to infer the epidemic parameters i.e. the probability
of transmission in each contact and the distribution of recovery times;
and (b) to infer missing data in a partial observation, e.g. correcting errors
or finding which of the two states S, R in the confused state setup.

The work is organized as follows. Section 2 provides a detailed description
of the graphical model representation of the stochastic epidemic dynamics.
The BP equations of the model and the details of their
efficient implementation are discussed in Section 3 (plus Appendices). The results of
the Bayesian inference under different observation models are reported in
Section 4. In Section 5, we present an efficient on-line method
for the inference of the epidemic parameters by maximization of the
log-likelihood by gradient ascent in the Bethe approximation. 

\section{Graphical model representation of the epidemic process}
We consider a discrete-time version of the  {\em susceptible-infected-recovered} (SIR) model \cite{kermack1932contributions} on a graph $G=(V,E)$ that represents the contact network of a set $V$ of individuals. A node $i$ can be in one of three possible states: susceptible $(S)$, infected $(I)$, and recovered/removed $(R)$. The state of node $i$ at time $t$ is represented by a variable $x_i^t \in \{S,I,R\}$. At each time step (e.g. a day) of the stochastic dynamics, an infected node $i$ can first spread the disease to each susceptible neighbor $j$ with given probability $\lambda_{ij}$, then recover with probability $\mu_i$. Once recovered, individuals do not get sick anymore.  
This process is Markovian, and satisfies $P(\mathbf{x}^{t+1}|\mathbf{x}^t)=\prod_i P(x_i^{t+1}|\mathbf{x}^t)$ where
\begin{eqnarray*} P(x_i^{t+1}=S|\mathbf {x}^t) &=& \mathbb{I}[x_i^t=S] \prod_{j\in\partial i} (1-\lambda_{ji} \mathbb{I}[x_j^t=I])\\ P(x_i^{t+1}=I|\mathbf {x}^t) &=& (1-\mu_i)\mathbb{I}[x_i^t=I] + \mathbb{I}[x_i^t=S] (1-\prod_{j\in\partial i} (1-\lambda_{ji} \mathbb{I}[x_j^t=I]))\\ P(x_i^{t+1}=R|\mathbf {x}^t) &=& \mu_i \mathbb{I}[x_i^t=I] + \mathbb{I}[x_i^t=R].\end{eqnarray*}

A realization of the stochastic dynamics is fully specified by knowing for each individual $i$, her {\em infection time} $t_i=\min \{t: x_i^t = I\}$ and her {\em recovery time} $g_i = \min\{g: x_i^{t_i+g+1} = R\}$. It is easy to show that, for a given initial configuration $\{x_i^0\}$, a realization of the stochastic process can be generated by drawing randomly the recovery time $g_{i}$ of each node $i$ and an infection {\em transmission delay} $s_{ij}$ from node $i$ to node $j$, for all pairs $(ij)$.
The recovery times $\{g_{i}\}$ are independent random variables extracted from geometric distributions
$\mathcal{G}_{i}\left(g_{i}\right)=\mu_{i}\left(1-\mu_{i}\right)^{g_{i}}$, while the delays $\{s_{ij}\}$ are also independent random variables distributed according to a truncated geometric distribution, 
\[
\omega_{ij}\left(s_{ij}|g_{i}\right)=\begin{cases}
\lambda_{ij}\left(1-\lambda_{ij}\right)^{s_{ij}}, & s_{ij}\leq g_{i}\\
\sum_{s>g_{i}}\lambda_{ij}\left(1-\lambda_{ij}\right)^{s}, & s_{ij}=\infty,
\end{cases}
\]
in which for convenience we concentrate in the value $s_{ij}=\infty$ the mass of the distribution beyond the hard cut-off $g_{i}$ imposed
by the recovery time. 
Infection times are related by the deterministic equation 
\begin{equation}
t_{i}=\min_{j\in\partial i} (t_{j}+s_{ji}) + 1 \label{eq:dynamical}
\end{equation}
which is a constraint encoding the infection dynamics of the SIR model. Then individual $i$ recovers at time $t_i + g_i$.
 
The exact mapping from realizations of the epidemic process to realizations of the transmission
delays and recovery times can be exploited to provide a graphical model representation of the
stochastic dynamics of the SIR model on a graph. For a given initial
condition, the joint probability distribution of infection
and recovery times conditioned on the initial state is
\begin{align}
\nonumber \mathcal{P}\left(\mathbf{t},\mathbf{g}|\mathbf{x}^{0}\right) & =\sum_{\mathbf{\{s_{ij}\}}}\mathcal{P}\left(\mathbf{s}|\mathbf{g}\right)\mathcal{P}\left(\mathbf{t}|\mathbf{x}^{0},\mathbf{g},\mathbf{s}\right)\mathcal{P\left(\mathbf{g}\right)} \\
& =\sum_{\mathbf{\{s_{ij}\}}}\prod_{i,j}\omega_{ij}\left(s_{ij}|g_{i}\right)\prod_{i}\phi_{i}(t_{i},\{t_k, s_{ki}\}_{k \in \partial i})\mathcal{G}_{i}(g_{i}),\label{eq:direct}
\end{align}
where 
\begin{equation}
	\phi_{i}(t_{i},\{t_k, s_{ki}\}_{k \in \partial i})=\delta(t_{i,}\mathbb{I}[x_{i}^{0}\neq I](\min_{k\in\partial i}(t_{k}+s_{ki})+1))
\end{equation}
is a characteristic function which imposes on each
node $i$ the dynamical constraint \eqref{eq:dynamical}.

In the following we derive a method to reconstruct information 
about the origin of the epidemics given some observation at later
times. We first need to compute the posterior probability of the
initial configuration given an observation at time $T$. This is
done by assuming to have a probabilistic prior on the initially infected
nodes and by applying Bayes formula,

\begin{equation}
\mathcal{P}\left(\mathbf{x}^{0}|\mathbf{x}^{T}\right)\propto\sum_{\mathbf{t,g}}\mathcal{P}\left(\mathbf{x}^{T}|\mathbf{t},\mathbf{g}\right)\mathcal{P}\left(\mathbf{t},\mathbf{g}|\mathbf{x}^{0}\right)\mathcal{P}\left(\mathbf{x}^{0}\right)=\sum_{\mathbf{t},\mathbf{g},\mathbf{s}}\prod_{i,j}\omega_{ij}\prod_{i}\phi_{i}\mathcal{G}_{i}\gamma_{i}\zeta_{i}\label{eq:bayes}
\end{equation}
where $\mathcal{P}\left(\mathbf{x}^{0}\right)=\prod_{i}\gamma_{i}\left(x_{i}^{0}\right)$
is a factorized prior on the initial infection and where we exploited
the fact that the state $\mathbf{x}^{T}$ is a deterministic function
of the set of infection and recovery times $(\mathbf{t},\mathbf{g})$
given by 

\begin{equation}
\mathcal{P}\left(\mathbf{x}^{T}|\mathbf{t},\mathbf{g}\right)=\prod_{i}\zeta_{i}^T\left(t_{i},g_{i},x_{i}^{T}\right)
\end{equation}
with 

\begin{equation}\label{zetait}
\zeta_{i}^t= \mathbb{I}\left[x_{i}^{t}=S, t < t_{i} \right] + \mathbb{I}\left[x_{i}^{t}=I,t_{i}\leq t<t_{i}+g_{i}\right]+\mathbb{I}\left[x_{i}^{t}=R,t_{i}+g_{i}\leq t\right].
\end{equation}

The above formula can be generalized to the case in which the parameters
$\mu$ and $\lambda$ have an explicit dependence on $t_{i}$. The
problem of computing the marginals from \eqref{eq:bayes} is in
general intractable (NP-hard) and we need to resort to an efficient
approximation. Here we choose to implement the BP
approximation which preserves some non trivial correlations between
variables and is exact in the limit cases in which correlation decay
holds. 

\begin{figure}[t]
\begin{center}
\includegraphics[width=12cm]{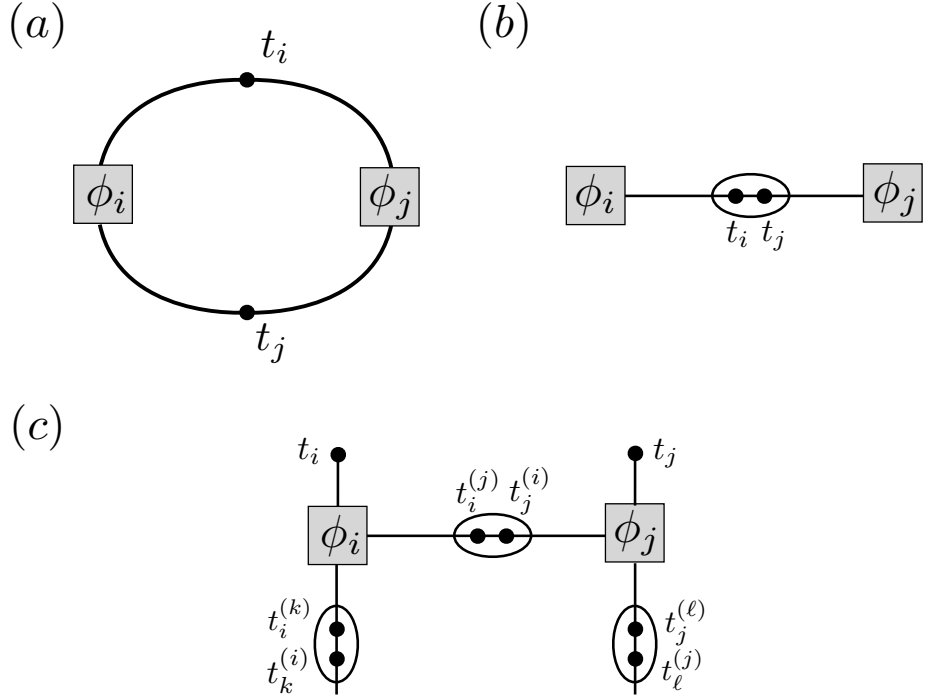}
\caption{(a) Example of a loopy factor graph representations induced by constraints such as those in  \eqref{eq:dynamical}. (b) Disentangled factor graph. (c) A more convenient representation of the disentangled factor graph employed in the present work. For simplicity, the dependency on $\{s_{ij}\}$ is not considered.}\label{fg0}
\end{center}
\end{figure}

We proceed by introducing a factor graph representation of \eqref{eq:bayes}, namely a bipartite
graph composed of factor nodes and variable nodes. In the standard definition, each variable appearing in the problem is identified by a variable node, while each factorized term of the probability weight in
\eqref{eq:bayes} is represented by a factor node. A factor node is connected to the set of 
 variable nodes appearing in the corresponding factorized term. However, with this definition the factor graph of \eqref{eq:bayes}
has a loopy structure both at local and global scale, and this could compromise the accuracy of the BP approximation. 
The existence of short loops can be easily verified even focusing only on the expression of the dynamical constraint \eqref{eq:dynamical}: a pair of variable nodes corresponding to the infection times $t_i$ and $t_j$ 
of neighboring individuals are indeed involved in the two factors $\phi_i$ and $\phi_j$, inducing a short loop in the factor graph (see Fig.\ref{fg0}a). 
We would like to use a factor graph representation that maintains the same topological properties of the original graph of contacts, in order to guarantee that BP is exact when the original graph of contacts is a tree. 
Following an approach proposed in \cite{altarelli_large_2013,altarelli_optimizing_2013}, the factor graph can be disentangled by grouping pairs of infection times $(t_i,t_j)$ in the same variable node as in Fig.\ref{fg0}b. For convenience,  we will keep all variable nodes $\{t_i\}$ but we will also introduce for each edge $(i,j)$ emerging from a node $i$ a set of copies $t_i^{(j)}$ of the infection time $t_i$, that will be forced to take the common value $t_i$ (see Fig.\ref{fg0}c) by including the constraint $\prod_{k\in\partial i}\delta(t_{i}^{(k)},t_{i})$ 
in the factor $\phi_i$.

We also observe that the factors $\phi_{i}$ depend on infection times and transmission delays just through the sums $t_{i}^{(j)}+s_{ij}$. 
It is thus more convenient to introduce the variables $t_{ij}=t_{i}^{(j)}+s_{ij}$
and express the dependencies through the pairs $(t_{i}^{(j)},t_{ij})$.

Finally it is convenient to group the variable $g_i$ with the corresponding infection times $t_i$ in the same variable node, replace $g_i$ and $g_j$ by their copies $g_i^{(j)}$ and $g_j^{(i)}$ in the edge constraints 
$\omega_{ij}(t_{ij}-t_{i}^{(j)}|g_{i}^{(i)})$ and $\omega_{ji}(t_{ji}-t_{j}|g_{j}^{(i)})$
and impose the identity $\prod_{k\in\partial i}\delta(g_{i}^{(k)},g_{i})$ for each node $i$.

We can now define the new factors

\begin{equation}
\phi_{ij}=\omega_{ij}(t_{ij} - t_{i}^{(j)}|g_{i}^{(i)})\omega_{ji}(t_{ji} - t_{j}^{(i)}|g_{j}^{(i)})
\end{equation} 

and

\begin{align}
	\nonumber \psi_{i}& =\delta(t_{i},\mathbb{I}[x_{i}^{0}\neq I](\min_{j\in\partial i}(t_{ji}+1)))\prod_{j\in\partial i}\delta(t_{i}^{(j)},t_{i})\text{\ensuremath{\delta}}(g_{i}^{(j)},g_{i}) \\
& =\phi_{i}(t_{i},\{{t}_{ji}\}_{j\in\partial i})\prod_{j\in\partial i}\delta(t_{i}^{(j)},t_{i})\text{\ensuremath{\delta}}(g_{i}^{(j)},g_{i}).
\end{align}

\begin{figure}[t]
\begin{center}
\includegraphics[width=12cm]{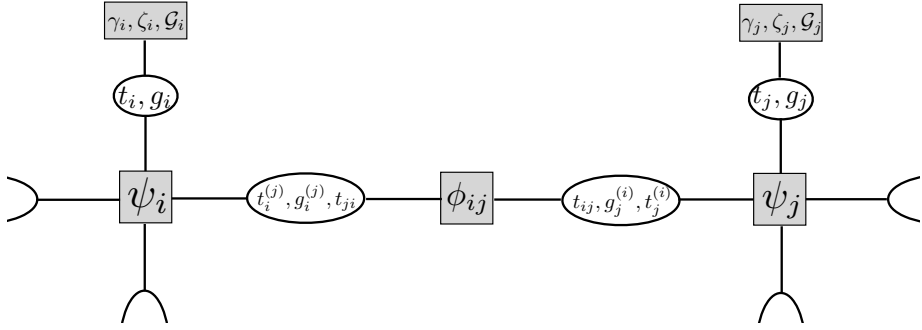}
\caption{Factor graph representation of the graphical model associated with the distribution \eqref{Qdist}.}\label{fg1}
\end{center}
\end{figure}

The posterior distribution can be written as 

\begin{equation}
	\mathcal{P}\left(\mathbf{x}^{0}|\mathbf{x}^{T}\right)\propto\sum_{\mathbf{{\bf }t},\{t_{ij}\},\mathbf{{\bf }g}}{\cal Q}({\bf g},{\bf t},\{t_{ij}\},x_{0})
\end{equation}

where

\begin{equation}\label{Qdist}
	{\cal Q}({\bf g},{\bf t},\{t_{ij}\},x_{0})=\frac{1}{Z}\prod_{i<j}\phi_{ij}\prod_{i}\psi_{i}{\cal G}_{i}\gamma_{i}\zeta_{i}.
\end{equation}

\begin{figure}[t]
\begin{center}
\includegraphics[width=12cm]{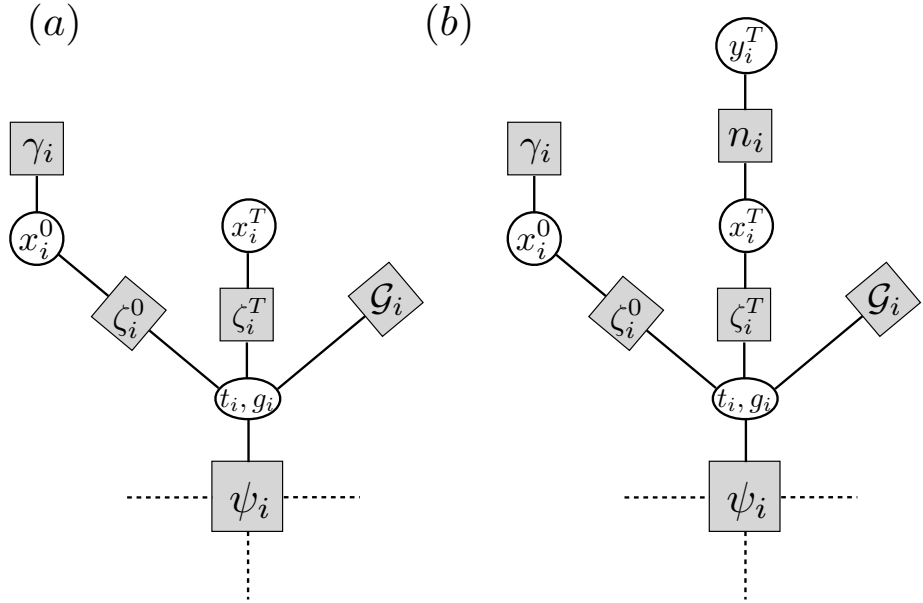}
\caption{(a) Detailed description of the internal structure of the factor node containing $\gamma_i,\zeta_i,\mathcal{G}_i$ in Fig.\ref{fg1}. (b) The modified factor graph used to include the observation models discussed in Section~\ref{section-observation}.}\label{fg2}
\end{center}
\end{figure}

Figure~\ref{fg1} shows the factor graph representation of the distribution \eqref{Qdist}. The factor node grouping $\xi_i,\zeta_i$ and $\mathcal{G}_i$ has in fact a more complex structure that is described in detail in Fig.\ref{fg2}a: the function $\zeta_i^t$, defined in \eqref{zetait}, connects the infection time $t_i$ with the state $x_i^t$ of the node at time $t$, $\gamma_i$ is the prior on the initial state, while $\mathcal{G}_i(g_i)$ is the recovery time distribution. Having built a factor graph with the same topology of the original graph, we can now compute the exact posterior marginals for the SIR model with BP in the case of tree graphs and obtain a good approximation of it even on graphs that are not trees if the correlation decay assumption is correct.
 
\section{BP Equations}

Belief propagation consists in a set of equations for
single-site probability distributions labeled by directed graph edges. These equations are solved by iteration, and on a fixed point give an approximation for single-site marginals and other quantities of interest like the partition function $Z$.

We recall the general form of the BP equations in the following. For a factorized probability measure on $\underline{z}=\{z_i\}$,
\begin{equation}
        M(\underline{z}) = \frac1Z \prod_a F_a(\underline{z}_a)
\end{equation}
where $\underline{z}_a$ is the subvector of variables that $F_a$ depends on, the general form of the equations is
\begin{eqnarray}
        p_{F_{a}\to i}\left(z_{i}\right) & = & \frac{1}{Z_{ai}} \sum_{\left\{ z_{j}:j\in \partial a\setminus i\right\} }F_{a}\left(\left\{ z_{i}\right\} _{i\in\partial a}\right)\prod_{j\in\partial a\setminus i}m_{j\to F_a}\left(z_{j}\right)\label{eq:factor-to-var}\\
        m_{i\to F_{a}}\left(z_{i}\right) & = & \frac{1}{Z_{ia}} \prod_{b\in \partial i\setminus a}p_{F_b\to i}\left(z_{i}\right)\label{eq:var-to-factor}\\
        m_{i}\left(z_{i}\right) & = & \frac{1}{Z_{i}} \prod_{b\in \partial i}p_{F_b\to i}\left(z_{i}\right)\label{eq:var-marginal}
\end{eqnarray}
where $F_{a}$ is a {\em factor} (i.e. $\psi_{i}$, $\phi_{ij}$, $\gamma_{i}$, $\zeta_{i}^0$, $\zeta_i^T$
or $\mathcal{G}_{i}$ in our case), $z_{i}$ is a variable (i.e.
$(t_{i},g_{i})$, $(t_{i}^{\left(j\right)},g_{i}^{\left(j\right)},t_{ji})$, $x_i^0$ or $x_i^T$
in our case), $\partial a$ is the subset of indices of variables in factor $F_a$ and $\partial i$ is the subset of factors that depend on $z_i$. The terms $Z_{ia},Z_{ai}$ and $Z_i$ are normalization factors that can be calculated once the rest of the right-hand side is computed. While equations \eqref{eq:var-to-factor}-\eqref{eq:var-marginal} can be always computed
efficiently in general, the computation of the trace in \eqref{eq:factor-to-var} may need a time which is exponential
in the number of participating variables. The update equations \eqref{eq:factor-to-var}
for factors $\phi_{ij}$, $\gamma_{i}$, $\zeta_{i}^0$, $\zeta_i^T$ and  $\mathcal{G}_{i}$ can be computed in a straightforward way because they involve a very small (constant) number of variables each.
In Appendix~\ref{appBP} we show the derivation of an efficient version of equation \eqref{eq:factor-to-var} for factor $\psi_{i}$ that can be computed in linear time in the degree of vertex $i$, and we provide a change of variables that simplifies the messages and further reduces the computation time of the update.

\section{Observation models}\label{section-observation}

In the inference of the origin of epidemic propagations it is often assumed that the state
of every node is known at the observation time $T$ with no uncertainty.
This is also the case studied in Ref.\cite{altarelli_prl_2014}, where the BP approach for this problem was first introduced.
In practice, every clinical test for determining the state of an
individual is affected by some amount of error, and this possibility
has to be taken into account in the inference problem. Therefore,
it is realistic to assume that each observation carries some level of
noise. We introduce a general concept of {\em observation model} for the inference problem, 
that allows dealing with several different cases of incomplete and noisy data using a common notation.
We will assume that the noise level is known (as it is the case for the majority
of clinical tests), and introduce a new variable ${y}^{T}_{i}\in\left\{ S,I,R\right\}$
for the observed state of node $i$ and an additional evidence
term that reflects the probability 
$n_{i}\left({y}^{T}_{i}| {x}^{T}_{i}\right)$ of the observed state ${y}^{T}_{i}$ given the true state ${x}^{T}_{i}$. In the factor graph, the
observed-state variables ${y}^{T}_{i}$ are fixed to their value given by the experimental observation (by means of a delta function representing an infinite external
field), while the true-state variables $x^{T}_{i}$ are
traced over in the compatibility function $\zeta_{i}^T$. More explicitly, the modified factor graph shown in Fig.~\ref{fg2}b, contains a 
$\zeta_{i}^T$ factor node attached to the true-state variable $x^{T}_{i}$, which is linked to the observed state ${y}^{T}_{i}$  (which is a constant) through the
node $n_{i}\left({y}^{T}_{i}|x^{T}_{i}\right)$. The posterior distribution now takes the form:

\begin{equation}
	\mathcal{P}\left(\mathbf{x}^{0}|\mathbf{y}^{T}\right)\propto\sum_{\mathbf{{\bf }x}^{T},\mathbf{{\bf }t},\mathbf{{\bf }t}_{ij},\mathbf{{\bf }g}}{\cal Q}'({\bf x}^{T},{\bf g},{\bf t},{\bf t}_{ij},x_{0})
\end{equation}

where

\begin{equation}\label{Qprime-dist}
	{\cal Q}'({\bf x}^{T},{\bf g},{\bf t},{\bf t}_{ij},x_{0})=\frac{1}{Z}\prod_{i<j}\phi_{ij}\prod_{i}\psi_{i}{\cal G}_{i}\gamma_{i}\zeta_{i}n_{i}.
\end{equation}

In what follows, we introduce for convenience a map~$\rho(s)$ from indices $i \in \{1,2,3\}$ into configurations of the $x$ variables, such that~$\rho(1)=S,~\rho(2)=I,~\rho(3)=R$, and then define the \emph{observational transition matrix} (\emph{OTM}) $O^{(i)}_{s,t}$ whose elements are the transition probabilities:

\begin{equation}
O^{(i)}_{s,t} = n_{i}\left(\rho(s),\rho(t)\right).
\end{equation}

The case in which observations are complete and noiseless corresponds to an identity
matrix $O^{(i)}_{s,t}=\delta_{st}$. In
the following sections, we will cover some interesting examples of
applications of this scheme to confused and noisy observations. Note that, in this generalized scheme, we can also take into account
the case of partial observations, by assuming a totally uniform \emph{OTM} $O^{(i)}_{s,t}\equiv\frac{1}{3}$ for 
unobserved nodes.

\subsection{Inference of epidemic source from confused observations}\label{section-confused}

In some situations, it could be hard to distinguish between nodes that already
recovered from a disease and nodes that did not contract it at all.
To take into account this fact, we follow the approach of Ref.\cite{zhu_information_2013}
and explore the efficiency of our inference machinery in a setting
in which observations on Susceptible and Recovered nodes are confused.
More specifically, we allow only two types of observed states $x_{i}^{T}\in\left\{ I,N\right\} $,
where $N$ stands for Not-Infected. This situation corresponds to
choose the following \emph{OTM}:

\[
O^{(i)}=\left(\begin{array}{ccc}
\frac{1}{2} & 0 & \frac{1}{2}\\
0 & 1 & 0\\
\frac{1}{2} & 0 & \frac{1}{2}
\end{array}\right).
\]

We verified the performances of the BP algorithm on a completely uniform setting provided by random regular
graphs with identical infection parameters $\left(\lambda,\mu\right)$ for all nodes and links.
All epidemic propagations were initiated from a unique seed (the zero
patient). For each node, the BP algorithm provides an estimate of
the posterior probability that the node got infected at a certain
time, and thus also the probability that the node was the origin of
the epidemics. We can thus rank the nodes in decreasing order with respect of the 
estimated probability of being the origin of the observed epidemics:
the position of the true origin in the ranking provided by the algorithm
is a good measure of the efficacy of the method. In what follows,
we indicate with $i_{0}$ the ranking of the true origin of the epidemics,
and with $|G|$ the number of nodes in the graph $G$.

An important by-product of the algorithm is the ability to infer the
true state of a node from the marginal of the infection time, providing in this way a method to
``correct'' observations. More precisely, in the present
example we consider the problem of discriminating between Susceptible
and Recovered nodes. An effective method for quantifying the accuracy
of such binary classification problem is the \emph{Receiver Operating
Characteristic} (\emph{ROC}) curve, namely a plot of the ``true positive
rate" against the ``false positive rate". Constructing the \emph{ROC}
curve in the present case is very easy: we select the $N$ nodes and
rank them on the base of their marginal $\mathcal{P}\left(t_{i} = \infty|\mathbf{x}^{0}\right)$,
we then take one step upward in the \emph{ROC} whenever a true positive
case is encountered (${y}_{i}^{T}=x_{i}^{T}=S$) or one step rightward in case of a 
false positive (${y}_{i}^{T}\neq x_{i}^{T}$). We performed this discrimination
analysis for each sample and then computed the average value of the
area under the \emph{ROC} curve, that gives indication of the fraction
of correctly classified nodes. It turns out that the proposed algorithm can
be effectively used as an ex-post-facto tool for discriminating Susceptible
against Recovered individuals.

\begin{figure}
\includegraphics[scale=0.8]{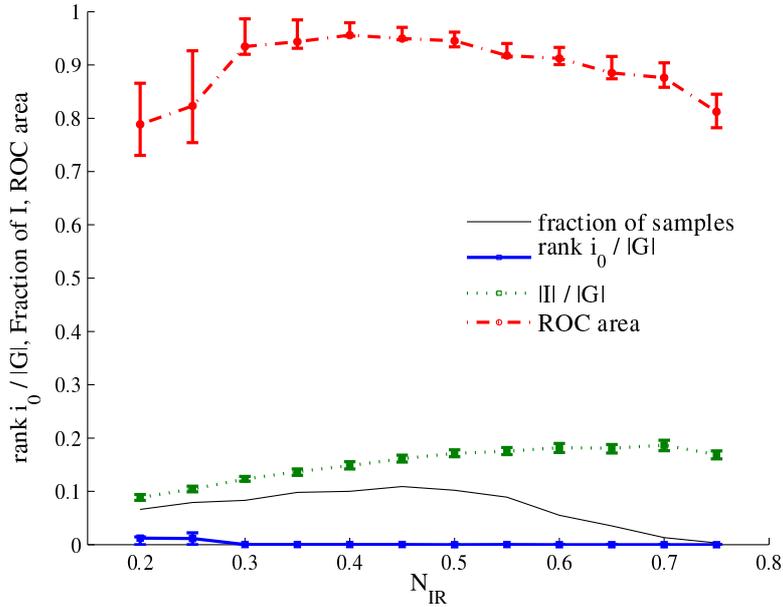} \protect\caption{\label{fig:rr10_T10_g1_l06_mu1_china} 
Average normalized rank of the true zero-patient (blue solid line), average ROC area (red dashed line) and average fraction of Infected nodes  $\frac{|I|}{|G|}$ (green dotted line)
as a function of the rescaled epidemic size $\frac{|I|+|R|}{|G|}$. The fraction of the $M$ samples belonging to each bin of the rescaled epidemic size is also indicated. 
The realization of the epidemic process is propagated for $T=10$ steps with $\lambda=0.6$ and $\mu=1$.
Observations are confused, i.e. $x_{i}^{t}\in\left\{ I,N\right\} $.
Simulations were run over $M=1000$ samples of random regular graphs
with $N=1000$ nodes and degree $d=4$.}
\end{figure}

Figure \ref{fig:rr10_T10_g1_l06_mu1_china} displays the average rank
of the true infected site $i_{0}$, normalized to the network size
$|G|$, for a set of $M=1000$ simulated epidemic propagations with
$\lambda=0.6$ and $\mu=1$ on random regular graphs of size $N=1000$
and degree $d=4$ ($T=10$). The quantities of interest are plot as functions of the normalized
epidemic size $N_{IR}=\frac{|I|+|R|}{|G|}$ (i.e. the fraction of infected
or recovered sites), whose values are discretized with intervals of width equal to $0.05$. Note that in all the figures we show in the paper, we discarded the rare cases with very low epidemic size ($N_{IR}<0.3$ in Fig. 5, $N_{IR}<0.2$ elsewhere) where the number of infected is extremely low and the inference is practically unfeasible.

For each set of data, the symbols report the mean value obtained averaging over the samples belonging to that 
interval and the error bars indicate the corresponding standard deviation. The average
fraction of Infected nodes and the fraction of samples in each bin
are reported as a reference. The normalized average rank of the true origin is very low for all values of the normalized
epidemic size, meaning that the algorithm is very effective in identifying the zero-patient.  We also show the average \emph{ROC} area,
which reveals that the inference algorithm allows a very good discrimination between $S$ and $R$
nodes.

\begin{figure}
\includegraphics[scale=0.8]{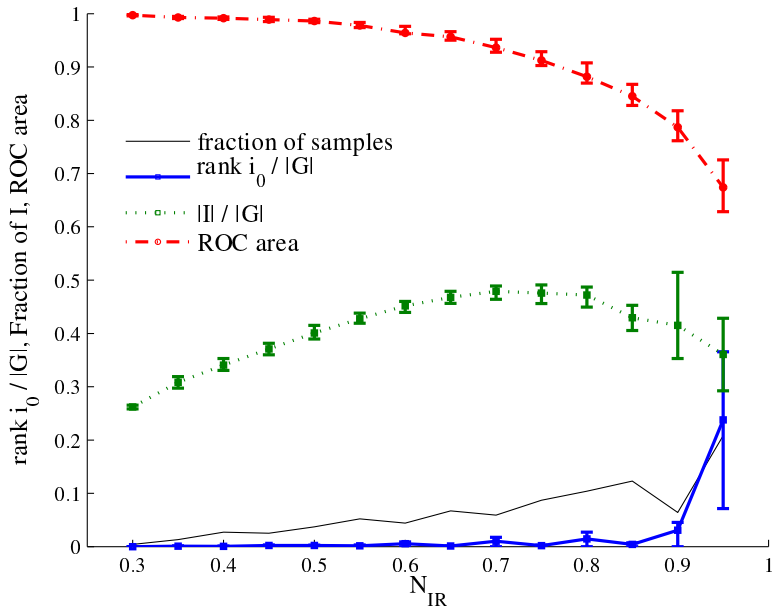}
\protect\caption{\label{fig:barab10_T7_g7_l06_mu05_china} Average normalized rank of the
true zero patient (blue solid line), average ROC area (red dashed line) and average fraction of Infected
nodes $\frac{|I|}{|G|}$ (green dotted line) as a function of the rescaled epidemic size $\frac{|I|+|R|}{|G|}$. 
The fraction of the $M$ samples belonging to each bin of the rescaled epidemic size is also indicated. 
The realization of the epidemic process is propagated up to the time step $T=7$ with $\lambda=0.6$ and $\mu=0.5$.
Observations are confused, i.e. $x_{i}^{t}\in\left\{ I,N\right\}$.
Simulations were run over $M=1000$ samples of Barabasi-Albert graphs
with $N=1000$ nodes and average degree $\hat{d}=4$.}
\end{figure}

The same analysis for a random graph with power-law degree distribution,
obtained using the Barabasi-Albert model \cite{Barabasi15101999},
is reported in Fig.\ref{fig:barab10_T7_g7_l06_mu05_china}. When the
observation time $T$ is sufficiently small ($T=7$ in Fig.\ref{fig:barab10_T7_g7_l06_mu05_china}),
the performance of the algorithm is high. When longer observation times are considered,
epidemics tend to cover the whole network and convergence issues emerge. In this regime, most of the infected nodes have already recovered at the observation time $T$ (and thus they cannot be distinguished anymore from the susceptible ones). This causes a rapid decay of the available information content that explains the performance degradation. A similar effect arises also on random regular graphs, but at longer times, as we will see in Section~\ref{section-noisy}.

In summary, even when supplied with confused observations, BP shows striking ability to discriminate between Recovered and
Susceptible nodes, provided that there is enough information at the chosen observation time $T$. 

\subsection{Inference of the epidemic source with noisy observations}\label{section-noisy}

Let us now consider a simple type of observational noise. Suppose that a node on state $x$, 
has probability $1-\nu$ of being correctly observed in state $x$, and probability $\nu$ of 
being observed incorrectly in one of the two remaining states, distributed uniformly among the two.
For example, node $i$ could be $I$ (Infected) at the observation time $T$, and, for a given noise level $\nu$, there will be an equal probability
$\frac{\nu}{2}$ for node $i$ to be observed in the $R$ (Recovered)
or $S$ (Susceptible) state. This setting corresponds to the following \emph{OTM}:
\[
O^{(i)}=\left(\begin{array}{ccc}
1-\nu & \frac{\nu}{2} & \frac{\nu}{2}\\
\frac{\nu}{2} & 1-\nu & \frac{\nu}{2}\\
\frac{\nu}{2} & \frac{\nu}{2} & 1-\nu
\end{array}\right).
\]

\begin{figure}
\includegraphics[scale=0.8]{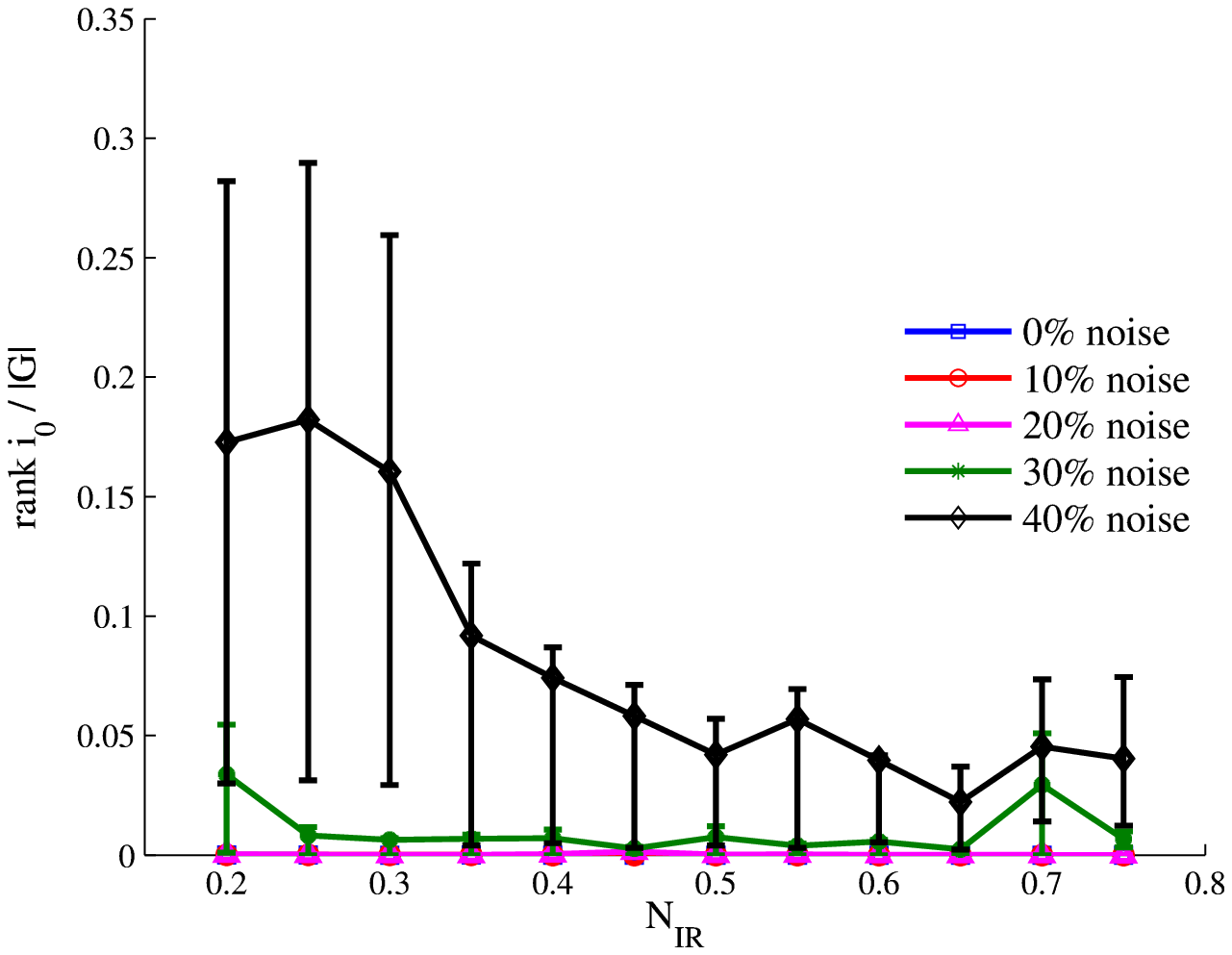} \protect\caption{\label{fig:rr_T10_lambda06_G10_mu06_noise}
Average normalized rank of the
true zero patient as a function of epidemic size $\frac{|I|+|R|}{|G|}$
for various levels of noise $\nu$ in the observation (the error-bars indicate the standard deviation computed on the sub-sample corresponding to a given epidemic size). 
Each curve refers to $M=1000$ samples of Random Regular graphs with $N=1000$ nodes
and degree $d=4$. Epidemics is propagated until $T=10$ with $\lambda=0.6$
and $\mu=1$.}
\end{figure}

\begin{figure}
\includegraphics[scale=0.8]{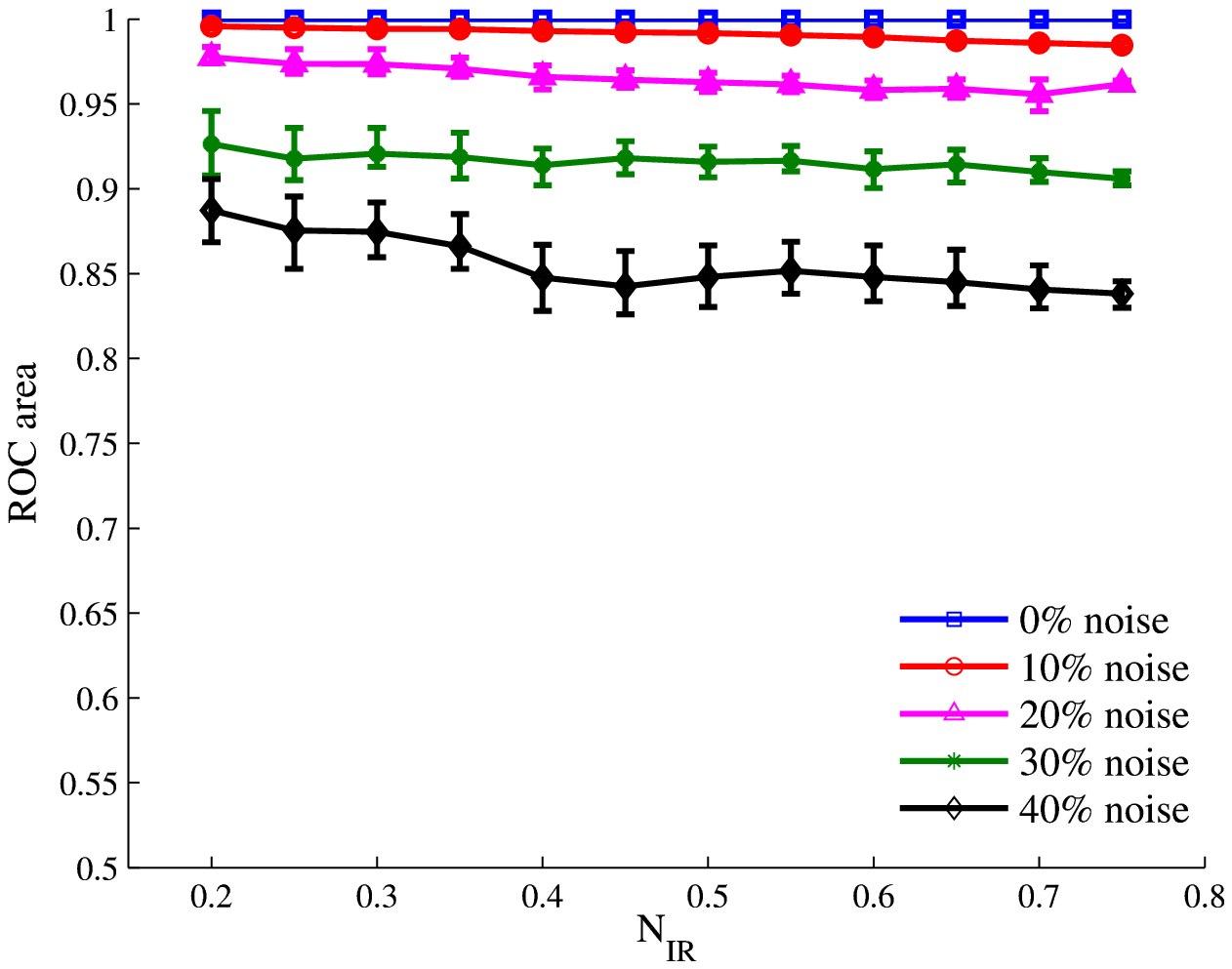}
\protect\caption{\label{fig:roc_rr_T10_lambda06_G10_mu06_noise} Average area of the
ROC curve as a function of epidemic size $\frac{|I|+|R|}{|G|}$ for
various levels of noise $\nu$ in the observation (the error-bars indicate the standard deviation computed on the sub-sample corresponding to a given epidemic size). 
Each curve refers to $M=1000$ samples of Random Regular graphs with $N=1000$ nodes and
degree $d=4$. Epidemics is propagated until $T=10$ with $\lambda=0.6$
and $\mu=1$.}
\end{figure}

We simulated a set of $M=1000$ single-source epidemic propagations
with $\lambda=0.6$ and $\mu=1$ on random regular graphs with $N=1000$
nodes and degree $d=4$. In Fig.~\ref{fig:rr_T10_lambda06_G10_mu06_noise}
we show the average rank of the true origin of the epidemics for various levels
of the observational noise up to $\nu=0.4$. The low values of the
average rank  obtained demonstrate that the BP
algorithm is able to perform extremely well up to very high levels
of noise. The corresponding ROC curves are plotted in Fig.~\ref{fig:roc_rr_T10_lambda06_G10_mu06_noise}.

\begin{figure}
\includegraphics[scale=0.7]{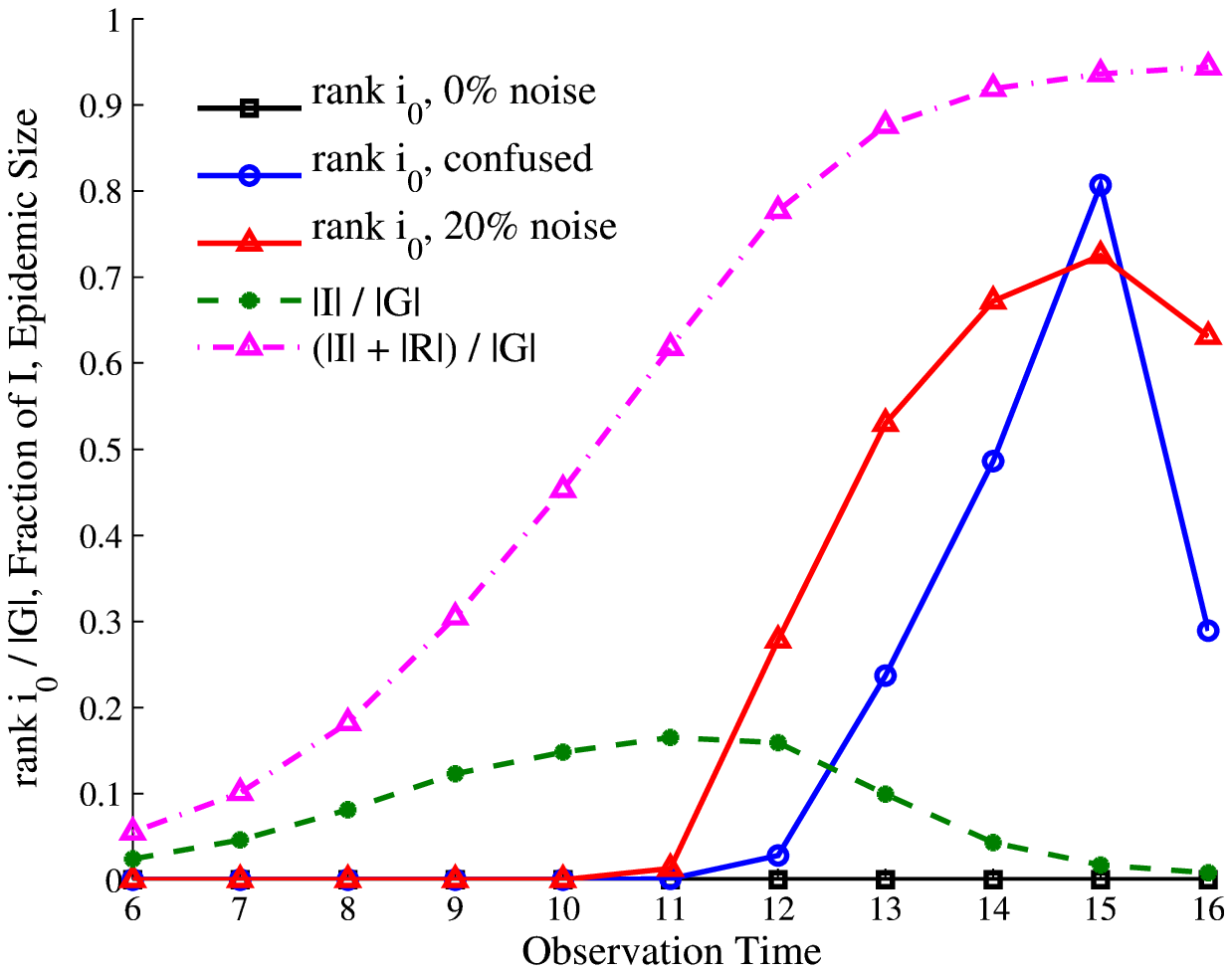}
\caption{\label{fig:on_time_rr_l06_d1} Normalized rank of the true zero-patient (solid lines), fraction of Infected nodes  $\frac{|I|}{|G|}$  (green dotted line) and rescaled epidemic size $\frac{|I|+|R|}{|G|}$ (dotted purple line) as a function of observation time $T$ for a single realization of the epidemic process, propagated with $\lambda=0.6$ and $\mu=1$, on  a random regular graph with $N=1000$ nodes and degree $d=4$.
Observations are complete (black solid line, superimposed to x axes), confused (blue solid line), and with noise $20\%$ noise level (red solid line).
Values of the normalized rank greater than $0.5$ are meaningless: they are the realization of a random variable with average close to 0.5, and they are evidence that for large $T$ the inference algorithm is unable to identify the zero-patient with better precision than pure chance.
}
\end{figure}

We investigated the role of observation time $T$ in relation to the amount of information needed for inferring the zero patient: simulations were run for given realizations of the epidemic process and observation time was systematically varied. In Fig.~\ref{fig:on_time_rr_l06_d1} we show a representative situation in random graphs (the picture is similar in scale-free graphs, as we argued in Section\ref{section-confused}). It turns out that the ratio of infected nodes to epidemic size is critical for inference: when observation time is too long so that the majority of infected individuals have recovered, it is much more difficult to find the zero-patient in the noisy and confused case. As it can be seen in the figure, this sharp change of behaviour (manifested at $T=11$) is present even in single instances. 

\section{Inference of Epidemic Parameters}

We have shown that if the parameters of the epidemics are known in
advance, our inference method can effectively detect the zero patient.
It is reasonable to assume that, for certain types of diseases, clinical
information could dictate some plausible range for the average rates of infection
and recovery. In the simplified case in which the infection parameters are
uniform among the population, the BP method can be generalized in a way to infer both the zero patient and the epidemic parameters 
at the same time.

\begin{figure}
\includegraphics[width=1\columnwidth]{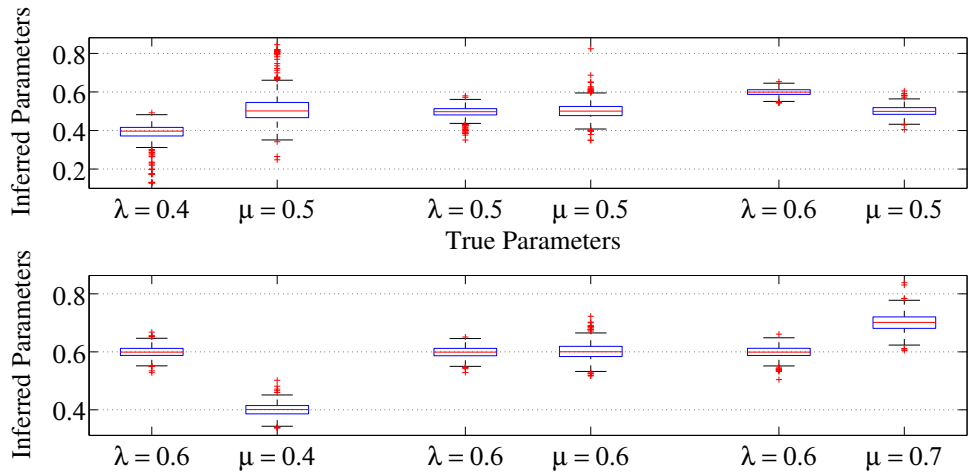} \protect\caption{\label{fig:rr_T10_G10_inference}Inferred epidemic parameters for
six different configurations of true $\left(\lambda,\mu\right)$ parameters.
Forward epidemic is simulated until observation time $T=10$. Each
pair of boxes refers to $M=1000$ instances of Random Regular graphs
with $N=1000$ nodes and degree $g=4$. Box edges signal the $25$th
and $75$th percentiles, the central red lines is the median. Whiskers
extend to cover $99.3\%$ of the data for a gaussian distribution.
Outliers are marked as red points outside the whiskers.}
\end{figure}

From our bayesian approach, $-f(\lambda,\mu)=\log Z(\lambda,\mu)$ is
the log-likelihood of the epidemic parameters for the observation $\mathbf{x}^{T}$.
Indeed, the log--likelihood of the parameters $-f(\lambda,\mu)$ equals
$\log\mathcal{P}\left(\mathbf{x}^{T}|\lambda,\mu\right)$ and the
latter can be computed as 
\begin{equation}
\mathcal{P}\left(\mathbf{x}^{T}|\lambda,\mu\right)=\sum_{\mathbf{t,g,x^{0}}}\mathcal{P}\left(\mathbf{x}^{T}|\mathbf{t},\mathbf{g}\right)\mathcal{P}\left(\mathbf{t},\mathbf{g}|\mathbf{x}^{0}\right)\mathcal{P}\left(\mathbf{x}^{0}\right)=Z\left(\lambda,\mu\right).
\end{equation}
In Ref.~\cite{altarelli_prl_2014}, this observation was used to infer the epidemic parameters
through an exaustive search in the space of parameters. This computation can be costly, wasting 
resources on noninteresting regions of the parameter space. Moreover, this type of experiments shows that 
the log-likelihood landscape with the Bethe approximation is generally very simple, presenting 
in most cases a single local maximum.
Here we describe a different method to
infer the  parameters together with the source of the epidemic outbreak.
The idea is to perform an on-line log-likelihood maximization through
gradient ascent in the Bethe approximation of the log-likelihood, by means of the
following updates: 
\begin{eqnarray}
\lambda & \leftarrow & \lambda+\epsilon\frac{\partial f}{\partial\lambda}\\
\mu & \leftarrow & \mu+\epsilon\frac{\partial f}{\partial\mu}
\end{eqnarray}
with $\epsilon$ a free convergence parameter. The free energy of the system can be approximated with the Bethe free energy, which in turn can
be expressed as a sum of local terms depending on BP messages. A detailed derivation of the expression of the derivatives $\frac{\partial f}{\partial\lambda}$ and $\frac{\partial f}{\partial\mu}$ of the Bethe free energy  is reported in Appendix~\ref{appGA}. 
In principle, the expressions obtained using the Bethe free energy
are valid only at the BP fixed point, and one should let BP updates
converge before making a step of gradient ascent. In practice,
we found that it is sufficient to interleave BP and gradient ascent
updates in order to obtain equivalent results. A fixed point of the 
interleaved updates is both a critical point of the Bethe log-likelihood and a BP approximation for the marginals.
We performed extensive simulations with a wide range of parameters
and found that, for reasonable fraction of infected nodes at the observation
time, the inference can simultaneously identify the zero patient perfectly and find good
estimates of the epidemic parameters. Some examples of inferred parameters
are shown in Fig.~\ref{fig:rr_T10_G10_inference} for six different
configurations of $\left(\lambda,\mu\right)$ parameters, with each
pair of box plots referring to $M=1000$ samples.

The method can be extended to treat the non-uniform case, at the expense of a higher computational
effort, that would amount in computing local derivatives of the free-energy function for each edge in the graph
with respect to edge-specific parameter. It should be clear that the number of parameters in the non-uniform case should not grow excessively
for inference purposes: we could, nevertheless, account for age or gender-dependent differences
in the probability to contract the disease or in the recovery rate with the introduction of additional information, attached to nodes and edges in the network.
Notice that, even in this case, an exhaustive search in the parameter space could be computationally too expensive.

\begin{figure}
\includegraphics[width=1\columnwidth]{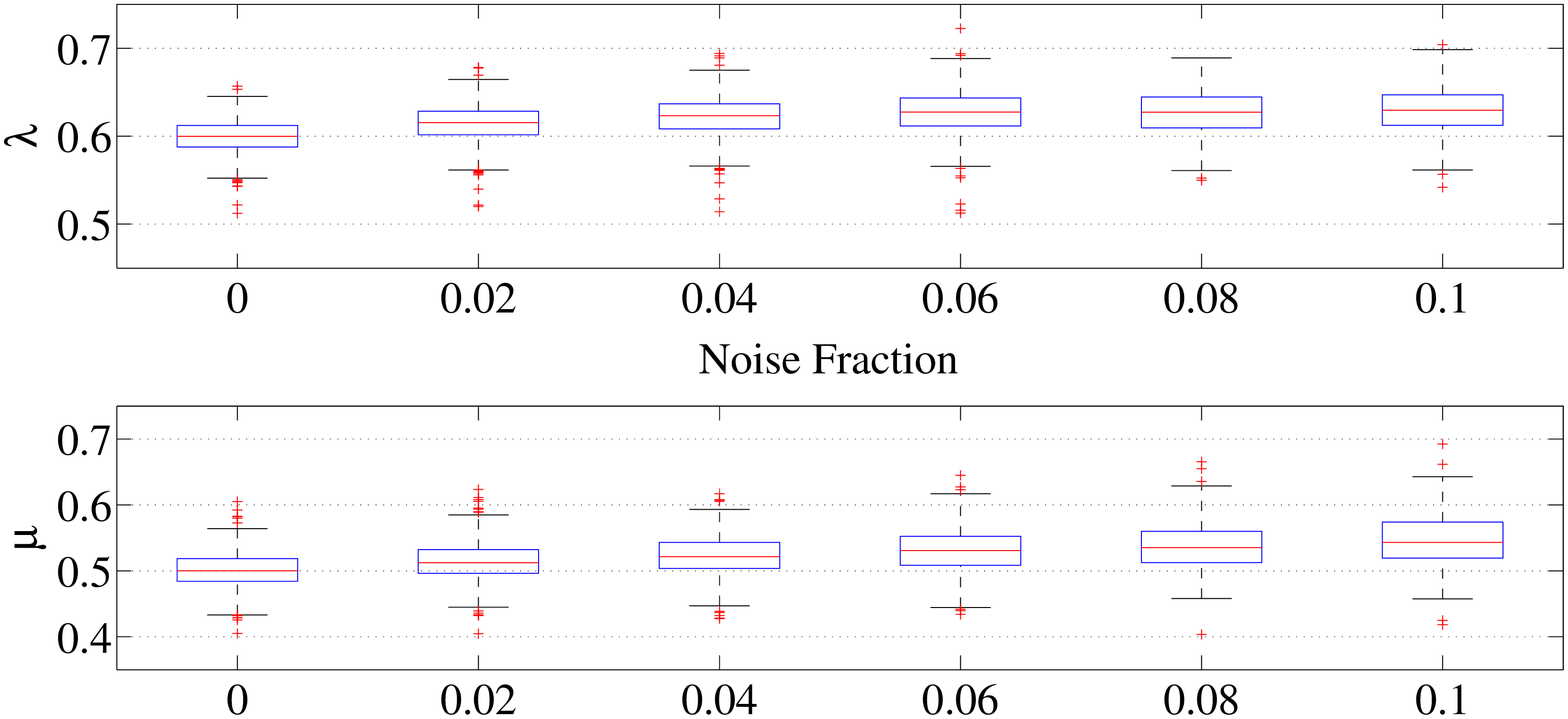}
\protect\caption{\label{fig:rr_T10_G10_inference_noise}Inferred epidemic parameters
for different observational noise rates $\nu$. Forward epidemic
is simulated until observation time $T=10$. Each box refers to $M=1000$
instances of Random Regular graphs with $N=1000$ nodes and degree
$g=4$. Box edges signal the $25$th and $75$th percentiles, the
central red lines is the median. Whiskers extend up to cover $99.3\%$
of the data for a gaussian distribution. Outliers are marked as red
points outside the whiskers.}
\end{figure}

The inference of parameters can be performed also in the presence of 
observational noise. In Fig.~\ref{fig:rr_T10_G10_inference_noise}
we show an example of inference for increasing levels of noise in
the observation, as defined in the preceeding section. Also in this
case the zero patient is detected with probability $1$ and the inferred
parameters are good estimators of the true values even up to a significant
fraction of noise.

We compared the ability of BP to infer the epidemic parameters with a simpler (though computationally expensive) procedure that does not require (nor provide) the simultaneous inference of the origin. The idea is to compare the statistical properties of the observation with the one of typical epidemics with given parameters $\lambda,\mu$, and choose those $\lambda,\mu$ that give properties that are closest (in a sense to be defined) to the ones observed.
More precisely, for each $\lambda\in\{0.05,0.1,\dots,0.95\}$ and $\mu \in \{0,0.05,\dots,1\}$, we generate 1000 random epidemics and compute the mean of the number of infected $I_{mean}(\lambda,\mu)$ and recovered $R_{mean}(\lambda,\mu)$ individuals. Afterwards, given an observation with $I$ infected and $R$ recovered individuals, we find 

$$(\lambda^*,\mu^*) = \arg\min_{\lambda,\mu} (I-I_{mean}(\lambda,\mu))^2 + (R-R_{mean}(\lambda,\mu))^2.$$

\begin{figure}
\includegraphics[width=0.45\columnwidth]{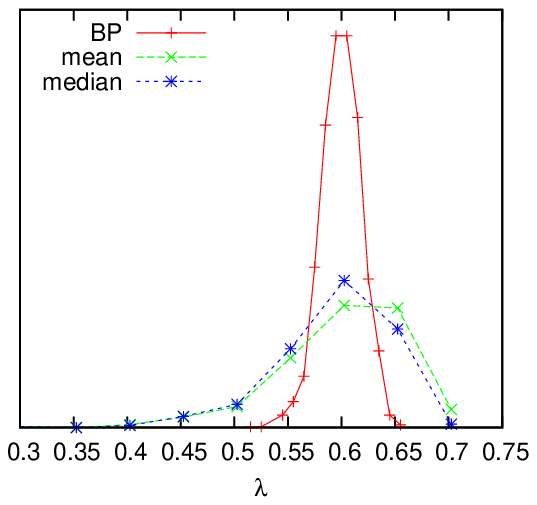}
\includegraphics[width=0.45\columnwidth]{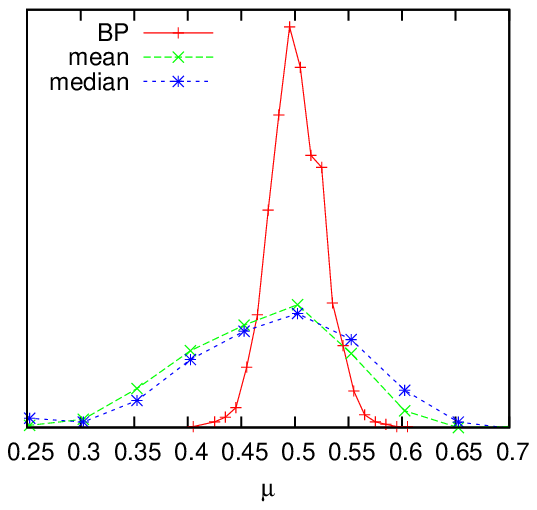}
\protect\caption{\label{fig:naive}Comparison of inference of epidemic parameters for $200$ random realizations with $\lambda=0.6$, $\mu=0.5$ between BP and the naive method consisting in finding the couple $(\lambda^*,\mu^*)$ which is closest in terms of mean (resp. median) number of infected and recovered individuals in euclidean distance. The distributions for the inference with BP correspond to the fifth example reported in Fig.\ref{fig:rr_T10_G10_inference} and the first in Fig.\ref{fig:rr_T10_G10_inference_noise}.}
\end{figure}

We also repeated the procedure using the $median$ instead of the 
mean (and computing thus $I_{median}$ and $R_{median}$). In Fig.~\ref{fig:naive} we show the distributions of $\lambda^*$ and $\mu^*$ found by the above procedure based on 200 epidemic realizations with $\lambda=0.6$ and $\mu=0.5$, along by the same distribution as found by the interleaved BP gradient ascent of the likelihood function. The results show that the BP-based procedure is able to infer the correct parameter $\lambda=0.6$ and $\mu = 0.5$ with much higher accuracy.

\section{Conclusion}
BP was recently proposed as an efficient tool for the inference of the origins of an epidemic propagation on graphs from a snapshot of the system at a later time. In the present work, we generalized the analysis to more realistic cases in which observations are imperfect. Experimental results show that BP performs well even in the presence of strong sources of uncertainty, such as observational noise, inability to distinguish between observed states, and uncertainty on the intrinsic model parameter. We provided
an exact solution on acyclic graphs for the first two problems, and a variational solution to compute the gradient of the likelihood of the parameters. The latter can be employed to find
local maxima of the likelihood function. We also characterized, by means of simulations, the amount of information that can be extracted on random graphs of various types, both on the past origin of the epidemics and on the missing bits on the present-time observation. Besides giving an excellent algorithmic answer to these questions, related to the past and the present of an observed epidemics, the scheme can be easily generalized to give accurate predictions about the future evolution of an outbreak from which only a partial observation (noisy and/or incomplete) of the current state is available. Work is in progress in this direction.

\appendix 

\section{Efficient BP updates}\label{appBP}

An efficient form for the update equations of the~$\psi_{i}$ factor nodes is the following:
\begin{eqnarray}
p_{\psi_{i}\to j}\left(t_{i}^{\left(j\right)},t_{ji},g_{i}^{\left(j\right)}\right) & \propto & \sum_{g_{i},t_{i}}\sum_{\left\{ t_{i}^{\left(k\right)},t_{ki},g_{i}^{\left(k\right)}\right\} }m_{i\to\psi_{i}}\left(t_{i},g_{i}\right)\times\\
 &  & \times\prod_{k\in\partial i\setminus j}m_{k\to\psi_{i}}\left(t_{i}^{\left(k\right)},t_{ki},g_{i}^{\left(k\right)}\right)\psi_{i}\left(t_{i},g_{i},\left\{ \left(t_{i}^{(k)},t_{ki},g_{i}^{(k)}\right)\right\} _{k\in\partial i}\right)\nonumber \\
 & \propto & m_{i\to\psi_{i}}\left(t_{i}^{\left(j\right)},g_{i}^{\left(j\right)}\right)\sum_{t_{ki}}\prod_{k\in\partial i\setminus j}m_{k\to\psi_{i}}\left(t_{i}^{\left(j\right)},t_{ki},g_{i}^{\left(j\right)}\right)\times\\
 &  & \times\left[\delta\left(t_{i}^{(j)},0\right)+\delta\left(t_{i}^{(j)},\left(1+\min_{k\in\partial i}\left\{ t_{ki}\right\} \right)\right)\right]\nonumber \\
 & \propto & \delta\left(t_{i}^{\left(j\right)},0\right)m_{i\to\psi_{i}}\left(0,g_{i}^{\left(j\right)}\right)\prod_{k\in\partial i\setminus j}\sum_{t_{ki}}m_{k\to\psi_{i}}\left(0,t_{ki},g_{i}^{\left(j\right)}\right)+\label{eq:pass}\\
 &  & +\; m_{i\to\psi_{i}}\left(t_{i}^{\left(j\right)},g_{i}^{\left(j\right)}\right)\I\left(t_{i}^{\left(j\right)}\leq t_{ji}+1\right)\prod_{k\in\partial i\setminus j}\sum_{t_{ki}\geq t_{i}^{\left(j\right)}-1}m_{k\to\psi_{i}}\left(t_{i}^{\left(j\right)},t_{ki},g_{i}^{\left(j\right)}\right)\nonumber \\
 &  & -\; m_{i\to\psi_{i}}\left(t_{i}^{\left(j\right)},g_{i}^{\left(j\right)}\right)\I\left(t_{i}^{\left(j\right)}<t_{ji}+1\right)\prod_{k\in\partial i\setminus j}\sum_{t_{ki}>t_{i}^{\left(j\right)}-1}m_{k\to\psi_{i}}\left(t_{i}^{\left(j\right)},t_{ki},g_{i}^{\left(j\right)}\right)\nonumber 
\end{eqnarray}
where in \eqref{eq:pass} we use the fact that 
\[
\delta\left(t_{i},\left(1+\min_{j\in\partial i}\left\{ t_{ji}\right\} \right)\right)=\prod_{j\in\partial i}\I\left(t_{i}\leq t_{ji}+1\right)-\prod_{j\in\partial i}\I\left(t_{i}<t_{ji}+1\right).
\] 

It is also possibile to use a simpler representation for the messages,
in which we just retain information on the relative timing of infection
time~$t_{i}^{\left(j\right)}$ for a node~$i$ and infection propagation
time~$t_{ji}$ on its link with node~$j$, introducing the variables
$\sigma_{ji}=1+\sign\left(t_{ji}-\left(t_{i}^{\left(j\right)}-1\right)\right)$.

In equation~\eqref{eq:pass} we can easily group the sums over different
configurations of~$\left(t_{ki},t_{i}^{(j)}\right)$ and write:
\begin{eqnarray}
p_{\psi_{i}\to j}\left(t_{i}^{\left(j\right)},\sigma_{ji},g_{i}^{\left(j\right)}\right) & \propto & \delta\left(t_{i}^{\left(j\right)},0\right)m_{i\to\psi_{i}}\left(0,g_{i}^{\left(j\right)}\right)\prod_{k\in\partial i\setminus j}\sum_{\sigma_{i}^{'\left(k\right)}}m_{k\to\psi_{i}}\left(0,\sigma_{ki},g_{i}^{\left(j\right)}\right)+\\
 &  & +\; m_{i\to\psi_{i}}\left(t_{i}^{\left(j\right)},g_{i}^{\left(j\right)}\right)\I\left(\sigma_{ji}=1,2\right)\prod_{k\in\partial i\setminus j}\sum_{\sigma_{ki}=1,2}m_{k\to\psi_{i}}\left(t_{i}^{\left(j\right)},\sigma_{ki},g_{i}^{\left(j\right)}\right)\nonumber \\
 &  & -\; m_{i\to\psi_{i}}\left(t_{i}^{\left(j\right)},g_{i}^{\left(j\right)}\right)\I\left(\sigma_{ji}=2\right)\prod_{k\in\partial i\setminus j}m_{k\to\psi_{i}}\left(t_{i}^{\left(j\right)},2,g_{i}^{\left(j\right)}\right)\nonumber 
\end{eqnarray}
Similarly, the outgoing message to the~$i$ variable node is: 
\begin{eqnarray}
p_{\psi_{i}\to i}\left(t_{i},g_{i}\right) & \propto & \delta\left(t_{i},0\right)\prod_{k\in\partial i}\sum_{\sigma_{ki}}m_{k\to\psi_{i}}\left(0,\sigma_{ki},g_{i}\right)+\\
 &  & +\prod_{k\in\partial i}\sum_{\sigma_{ki}=1,2}m_{k\to\psi_{i}}\left(t_{i},\sigma_{ki},g_{i}\right)\nonumber \\
 &  & -\prod_{k\in\partial i}m_{k\to\psi_{i}}\left(t_{i},2,g_{i}\right)\nonumber 
\end{eqnarray}
In the simplified ~$\left(t,\sigma,g\right)$ representation for
the messages, the update equation for the~$\phi_{ij}$ nodes can
be written in an easy way: 
\begin{equation}
p_{\phi_{ij}\to j}\left(t_{j},\sigma_{ij},g_{j}\right)\propto\sum_{t_{i},\sigma_{ji},g_{i}}\Omega\left(t_{i},t_{j},\sigma_{ij},\sigma_{ji},g_{i},g_{j}\right)p_{i\to\phi_{ij}}\left(t_{i},\sigma_{ji},g_{i}\right)
\end{equation}
where: 
\begin{equation}
\Omega\left(t_{i},t_{j},\sigma_{ij},\sigma_{ji},g_{i},g_{j}\right)=\begin{cases}
\chi\left(t_{i},t_{j},\sigma_{ij},g_{i}\right) & :t_{i}<t_{j},~\sigma_{ji}=2,~\sigma_{ij}\neq2\\
\chi\left(t_{i},t_{j},\sigma_{ij},g_{i}\right)+\left(1-\lambda\right)^{g_{i}+1} & :t_{i}<t_{j},~\sigma_{ji}=2,~\sigma_{ji}=2\\
\chi\left(t_{j},t_{i},\sigma_{ji},g_{j}\right) & :t_{j}<t_{i},~\sigma_{ji}=2,~\sigma_{ji}\neq2\\
\chi\left(t_{j},t_{i},\sigma_{ji},g_{j}\right)+\left(1-\lambda\right)^{g_{j}+1} & :t_{j}<t_{i},~\sigma_{ij}=2,~\sigma_{ji}=2\\
1 & :t_{i}=t_{j},~\sigma_{ji}=\sigma_{ij}=2\\
0 & :otherwise
\end{cases}\label{eq:omega_phi_sigma}
\end{equation}
and 
\begin{equation}
\chi\left(t_{1},t_{2},\sigma,g\right)=\sum_{t=t_{1}}^{t_{1}+g}\delta\left(\sigma\left(t_{2},t\right),\sigma\right)\lambda\left(1-\lambda\right)^{t-t_{1}}
\end{equation}
Note that it is possible to exploit the symmetry in $i$ and $j$
between the rows 1-2 and rows 3-4 in the definition~\eqref{eq:omega_phi_sigma}
for~$\Omega$: when one loops over the variables~$\left(t_{j},\sigma_{ij},g_{j}\right)$
in order to fill in the output message $p_{\phi_{ij}\to j}\left(t_{j},\sigma_{ij},g_{j}\right)$,
this consists in looping over the variables $\left(t_{j},\sigma_{ij},g_{j}\right)$
of the input message $p_{i\to\phi_{ij}}\left(t_{i},\sigma_{i},g_{i}\right)$,
just with a switch in indices. In the implementation, this
amounts in a significant reduction of a factor~$G$ in the computational
complexity for updates involving the factor node~$\phi_{ij}$.

\section{Gradient ascent method for the inference of the epidemic parameters}\label{appGA}

The free energy of the system can be approximated with the Bethe free energy that can
be expressed as a sum of local terms depending on the BP messages.
\begin{equation}
-f=\sum_{a}f_{a}+\sum_{i}f_{i}-\sum_{(ia)}f_{(ia)}\label{eq:free-energy}
\end{equation}
where 
\begin{eqnarray}
f_{a} & = & \log\left(\sum_{\left\{ z_{i}:i\in\dd a\right\} }F_{a}\left(\left\{ z_{i}\right\} _{i\in\dd a}\right)\prod_{i\in\dd a}m_{i\to a}(z_{i})\right)\\
f_{(ia)} & = & \log\left(\sum_{z_{i}}m_{i\to a}(z_{i})p_{F_{a}\to i}(z_{i})\right)\\
f_{i} & = & \log\left(\sum_{z_{i}}\prod_{b\in\dd i}p_{F_{b}\to i}(z_{i})\right)
\end{eqnarray}
The computation of the gradient of the free energy deserves some special
attention: being $f$ a function of all the BP messages, one would
argue that this messages depend on the model parameters too, at every
step in the BP algorithm. Actually, there is no need to consider this
implicit $(\lambda,\mu)$ dependence if BP has reached its fixed point,
that is when BP equations are satisfied and the messages are nothing
else but Lagrange multipliers with respect to the constraint minimization
of the Bethe free energy functional \cite{yedidia2001bethe}. In our
scheme, the only explicit dependence of free energy on epidemic parameters
is in the factor node terms $f_{a}$'s involving compatibility functions
$\phi_{ij}=\omega_{ij}\left(t_{ij}-t_{i}|g_{i}\right)\omega_{ji}\left(t_{ji}-t_{j}|g_{j}\right)$
and $\mathcal{G}_{i}\left(g_{i}\right)=\mu_{i}\left(1-\mu_{i}\right)^{g_{i}}$,
and the gradient can be computed very easily. For the $\phi_{ij}$
nodes we have: 
\begin{equation}
\frac{\partial f_{\phi_{ij}}}{\partial\lambda}=\frac{\sum_{t_{i},t_{ji},g_{i},t_{j},t_{ij},g_{j}}\frac{\partial\phi_{ij}}{\partial\lambda}\left(t_{i},t_{ji},g_{i},t_{j},t_{ij},g_{j}\right)m_{i\to\phi_{ij}}\left(t_{i},t_{ji},g_{i}\right)m_{j\to\phi_{ij}}\left(t_{j},t_{ij},g_{j}\right)}{\sum_{t_{i},t_{ji},g_{i},t_{j},t_{ij},g_{j}}\phi_{ij}\left(t_{i},t_{ji},g_{i},t_{j},t_{ij},g_{j}\right)m_{i\to\phi_{ij}}\left(t_{i},t_{ji},g_{i}\right)m_{j\to\phi_{ij}}\left(t_{j},t_{ij},g_{j}\right)}
\end{equation}
where 
\begin{equation}
\frac{\partial\phi_{ij}}{\partial\lambda}=\begin{cases}
1 & t_{i}<t_{j}\mbox{ and }t_{i}=t_{ij}<t_{i}+g_{i}\\
-\left(g_{i}-t_{i}\right)\lambda\left(1-\lambda\right)^{g_{i}-t_{i}-1} & t_{i}<t_{j}\mbox{ and }t_{i}<t_{ij}=t_{i}+g_{i}\\
\left(1-\lambda\right)^{t_{ij}-t_{i}}-\left(t_{ij}-t_{i}\right)\lambda\left(1-\lambda\right)^{t_{ij}-t_{i}-1} & t_{i}<t_{j}\mbox{ and }t_{i}<t_{ij}<t_{i}+g_{i}\\
1 & t_{j}<t_{i}\mbox{ and }t_{j}=t_{j}<t_{j}+g_{j}\\
-\left(g_{j}-t_{j}\right)\lambda\left(1-\lambda\right)^{g_{j}-t_{j}-1} & t_{j}<t_{i}\mbox{ and }t_{j}<t_{ji}=t_{j}+g_{j}\\
\left(1-\lambda\right)^{t_{ji}-t_{j}}-\left(t_{ji}-t_{j}\right)\lambda\left(1-\lambda\right)^{t_{ji}-t_{j}-1} & t_{j}<t_{i}\mbox{ and }t_{j}<t_{ji}<t_{j}+g_{j}\\
0 & \mbox{else}
\end{cases}\label{eq:phi_ij_free_energy_tprime}
\end{equation}
In the simplified~$\left(t,\sigma,g\right)$ representation for the
messages, equation~\eqref{eq:phi_ij_free_energy_tprime} takes the
form: 
\begin{equation}
\frac{\partial\phi_{ij}}{\partial\lambda}=\begin{cases}
\chi\left(t_{i},t_{j},\sigma_{ij},g_{i}\right) & t_{i}<t_{j},\sigma_{ji}=2,\sigma_{ij}\neq2\\
\chi\left(t_{i},t_{j},\sigma_{ij},g_{i}\right)-\left(g_{i}+1\right)\left(1-\lambda\right)^{g_{i}} & t_{i}<t_{j},\sigma_{ji}=2,\sigma_{ij}=2\\
\chi\left(t_{j},t_{i},\sigma_{ji},g_{j}\right) & t_{j}<t_{i},\sigma_{ji}=2,\sigma_{ij}\neq2\\
\chi\left(t_{j},t_{i},\sigma_{ji},g_{j}\right)-\left(g_{j}+1\right)\left(1-\lambda\right)^{g_{j}} & t_{j}<t_{i},\sigma_{ji}=2,\sigma_{ij}=2\\
0 & otherwise
\end{cases}\label{eq:phi_ij_free_energy_sigma}
\end{equation}
where: 
\begin{equation}
\chi\left(t_{1},t_{2},\sigma,g\right)=\sum_{t=t_{1}}^{t_{1}+g}\delta\left(\sigma\left(t_{2},t\right),\sigma\right)\left(1-\lambda\right)^{t-t_{1}}-\left(t-t_{1}\right)\lambda\left(1-\lambda\right)^{t-t_{1}-1}
\end{equation}
For the $\mathcal{G}_{i}$ nodes we have: 
\begin{equation}
\frac{\partial f_{\mathcal{G}_{i}}}{\partial\mu}=\frac{\sum_{g_{i}}\mathcal{\tilde{G}}_{i}(g_{i})m_{i\to\mathcal{G}_{i}}(g_{i})}{\sum_{g_{i}}\mathcal{G}_{i}(g_{i})m_{i\to\mathcal{G}_{i}}(g_{i})}
\end{equation}
where 
\begin{equation}
\mathcal{\tilde{G}}_{i}(g_{i})=\begin{cases}
\left(1-\mu\right)^{g_{i}}-g_{i}\mu\left(1-\mu\right)^{g_{i}-1} & :g_{i}<G\\
G-G\left(1-\mu\right)^{G-1} & :g_{i}=G.
\end{cases}
\end{equation}


\begin{thebibliography}{26}%
\makeatletter
\providecommand \@ifxundefined [1]{%
 \@ifx{#1\undefined}
}%
\providecommand \@ifnum [1]{%
 \ifnum #1\expandafter \@firstoftwo
 \else \expandafter \@secondoftwo
 \fi
}%
\providecommand \@ifx [1]{%
 \ifx #1\expandafter \@firstoftwo
 \else \expandafter \@secondoftwo
 \fi
}%
\providecommand \natexlab [1]{#1}%
\providecommand \enquote  [1]{``#1''}%
\providecommand \bibnamefont  [1]{#1}%
\providecommand \bibfnamefont [1]{#1}%
\providecommand \citenamefont [1]{#1}%
\providecommand \href@noop [0]{\@secondoftwo}%
\providecommand \href [0]{\begingroup \@sanitize@url \@href}%
\providecommand \@href[1]{\@@startlink{#1}\@@href}%
\providecommand \@@href[1]{\endgroup#1\@@endlink}%
\providecommand \@sanitize@url [0]{\catcode `\\12\catcode `\$12\catcode
  `\&12\catcode `\#12\catcode `\^12\catcode `\_12\catcode `\%12\relax}%
\providecommand \@@startlink[1]{}%
\providecommand \@@endlink[0]{}%
\providecommand \url  [0]{\begingroup\@sanitize@url \@url }%
\providecommand \@url [1]{\endgroup\@href {#1}{\urlprefix }}%
\providecommand \urlprefix  [0]{URL }%
\providecommand \Eprint [0]{\href }%
\providecommand \doibase [0]{http://dx.doi.org/}%
\providecommand \selectlanguage [0]{\@gobble}%
\providecommand \bibinfo  [0]{\@secondoftwo}%
\providecommand \bibfield  [0]{\@secondoftwo}%
\providecommand \translation [1]{[#1]}%
\providecommand \BibitemOpen [0]{}%
\providecommand \bibitemStop [0]{}%
\providecommand \bibitemNoStop [0]{.\EOS\space}%
\providecommand \EOS [0]{\spacefactor3000\relax}%
\providecommand \BibitemShut  [1]{\csname bibitem#1\endcsname}%
\let\auto@bib@innerbib\@empty
\bibitem [{\citenamefont {Bailey}(1975)}]{bailey_mathematical_1975}%
  \BibitemOpen
  \bibfield  {author} {\bibinfo {author} {\bibfnamefont {N.~T.~J.}\
  \bibnamefont {Bailey}},\ }\href@noop {} {\emph {\bibinfo {title} {The
  mathematical theory of infectious diseases and its applications}}}\ (\bibinfo
   {publisher} {Griffin},\ \bibinfo {address} {London},\ \bibinfo {year}
  {1975})\BibitemShut {NoStop}%
\bibitem [{\citenamefont {Kermack}\ and\ \citenamefont
  {McKendrick}(1927)}]{kermack1927contribution}%
  \BibitemOpen
  \bibfield  {author} {\bibinfo {author} {\bibfnamefont {W.}~\bibnamefont
  {Kermack}}\ and\ \bibinfo {author} {\bibfnamefont {A.}~\bibnamefont
  {McKendrick}},\ }\bibfield  {title} {\enquote {\bibinfo {title} {A
  contribution to the mathematical theory of epidemics},}\ }\href@noop {}
  {\bibfield  {journal} {\bibinfo  {journal} {Proceedings of the Royal Society
  of London. Series A}\ }\textbf {\bibinfo {volume} {115}},\ \bibinfo {pages}
  {700} (\bibinfo {year} {1927})}\BibitemShut {NoStop}%
\bibitem [{\citenamefont {Isella}\ \emph {et~al.}(2011)\citenamefont {Isella},
  \citenamefont {Stehl{\'e}}, \citenamefont {Barrat}, \citenamefont {Cattuto},
  \citenamefont {Pinton},\ and\ \citenamefont {Van~den
  Broeck}}]{isella_whats_2011}%
  \BibitemOpen
  \bibfield  {author} {\bibinfo {author} {\bibfnamefont {L.}~\bibnamefont
  {Isella}}, \bibinfo {author} {\bibfnamefont {J.}~\bibnamefont {Stehl{\'e}}},
  \bibinfo {author} {\bibfnamefont {A.}~\bibnamefont {Barrat}}, \bibinfo
  {author} {\bibfnamefont {C.}~\bibnamefont {Cattuto}}, \bibinfo {author}
  {\bibfnamefont {J.-F.}\ \bibnamefont {Pinton}}, \ and\ \bibinfo {author}
  {\bibfnamefont {W.}~\bibnamefont {Van~den Broeck}},\ }\bibfield  {title}
  {\enquote {\bibinfo {title} {What's in a crowd? analysis of face-to-face
  behavioral networks},}\ }\href {\doibase 10.1016/j.jtbi.2010.11.033}
  {\bibfield  {journal} {\bibinfo  {journal} {Journal of Theoretical Biology}\
  }\textbf {\bibinfo {volume} {271}},\ \bibinfo {pages} {166} (\bibinfo {year}
  {2011})}\BibitemShut {NoStop}%
\bibitem [{\citenamefont {Rocha}\ \emph {et~al.}(2010)\citenamefont {Rocha},
  \citenamefont {Liljeros},\ and\ \citenamefont
  {Holme}}]{rocha_information_2010}%
  \BibitemOpen
  \bibfield  {author} {\bibinfo {author} {\bibfnamefont {L.~E.~C.}\
  \bibnamefont {Rocha}}, \bibinfo {author} {\bibfnamefont {F.}~\bibnamefont
  {Liljeros}}, \ and\ \bibinfo {author} {\bibfnamefont {P.}~\bibnamefont
  {Holme}},\ }\bibfield  {title} {\enquote {\bibinfo {title} {Information
  dynamics shape the sexual networks of internet-mediated prostitution},}\
  }\href {\doibase 10.1073/pnas.0914080107} {\bibfield  {journal} {\bibinfo
  {journal} {{PNAS}}\ }\textbf {\bibinfo {volume} {107}},\ \bibinfo {pages}
  {5706} (\bibinfo {year} {2010})},\ \bibinfo {note} {{PMID:}
  20231480}\BibitemShut {NoStop}%
\bibitem [{\citenamefont {Danon}\ \emph {et~al.}(2011)\citenamefont {Danon},
  \citenamefont {Ford}, \citenamefont {House}, \citenamefont {Jewell},
  \citenamefont {Keeling}, \citenamefont {Roberts}, \citenamefont {Ross},\ and\
  \citenamefont {Vernon}}]{danon2011networks}%
  \BibitemOpen
  \bibfield  {author} {\bibinfo {author} {\bibfnamefont {L.}~\bibnamefont
  {Danon}}, \bibinfo {author} {\bibfnamefont {A.~P.}\ \bibnamefont {Ford}},
  \bibinfo {author} {\bibfnamefont {T.}~\bibnamefont {House}}, \bibinfo
  {author} {\bibfnamefont {C.~P.}\ \bibnamefont {Jewell}}, \bibinfo {author}
  {\bibfnamefont {M.~J.}\ \bibnamefont {Keeling}}, \bibinfo {author}
  {\bibfnamefont {G.~O.}\ \bibnamefont {Roberts}}, \bibinfo {author}
  {\bibfnamefont {J.~V.}\ \bibnamefont {Ross}}, \ and\ \bibinfo {author}
  {\bibfnamefont {M.~C.}\ \bibnamefont {Vernon}},\ }\bibfield  {title}
  {\enquote {\bibinfo {title} {Networks and the epidemiology of infectious
  disease},}\ }\href@noop {} {\bibfield  {journal} {\bibinfo  {journal}
  {Interdisciplinary perspectives on infectious diseases}\ }\textbf {\bibinfo
  {volume} {2011}} (\bibinfo {year} {2011})}\BibitemShut {NoStop}%
\bibitem [{\citenamefont {Marathe}\ and\ \citenamefont
  {Vullikanti}(2013)}]{marathe_2013}%
  \BibitemOpen
  \bibfield  {author} {\bibinfo {author} {\bibfnamefont {M.}~\bibnamefont
  {Marathe}}\ and\ \bibinfo {author} {\bibfnamefont {A.~K.~S.}\ \bibnamefont
  {Vullikanti}},\ }\bibfield  {title} {\enquote {\bibinfo {title}
  {Computational epidemiology},}\ }\href {\doibase 10.1145/2483852.2483871}
  {\bibfield  {journal} {\bibinfo  {journal} {Commun. ACM}\ }\textbf {\bibinfo
  {volume} {56}},\ \bibinfo {pages} {88} (\bibinfo {year} {2013})}\BibitemShut
  {NoStop}%
\bibitem [{\citenamefont {Shah}\ and\ \citenamefont
  {Zaman}(2010)}]{shah_detecting_2010}%
  \BibitemOpen
  \bibfield  {author} {\bibinfo {author} {\bibfnamefont {D.}~\bibnamefont
  {Shah}}\ and\ \bibinfo {author} {\bibfnamefont {T.}~\bibnamefont {Zaman}},\
  }\bibfield  {title} {\enquote {\bibinfo {title} {Detecting sources of
  computer viruses in networks: theory and experiment},}\ }\href@noop {}
  {\bibfield  {journal} {\bibinfo  {journal} {{SIGMETRICS}}\ }\textbf {\bibinfo
  {volume} {38}},\ \bibinfo {pages} {203{\textendash}214} (\bibinfo {year}
  {2010})}\BibitemShut {NoStop}%
\bibitem [{\citenamefont {Comin}\ and\ \citenamefont
  {da~Fontoura~Costa}(2011)}]{comin_identifying_2011}%
  \BibitemOpen
  \bibfield  {author} {\bibinfo {author} {\bibfnamefont {C.~H.}\ \bibnamefont
  {Comin}}\ and\ \bibinfo {author} {\bibfnamefont {L.}~\bibnamefont
  {da~Fontoura~Costa}},\ }\bibfield  {title} {\enquote {\bibinfo {title}
  {Identifying the starting point of a spreading process in complex
  networks},}\ }\href {\doibase 10.1103/PhysRevE.84.056105} {\bibfield
  {journal} {\bibinfo  {journal} {Phys. Rev. E}\ }\textbf {\bibinfo {volume}
  {84}},\ \bibinfo {pages} {056105} (\bibinfo {year} {2011})}\BibitemShut
  {NoStop}%
\bibitem [{\citenamefont {Shah}\ and\ \citenamefont
  {Zaman}(2011)}]{shah_rumors_2011}%
  \BibitemOpen
  \bibfield  {author} {\bibinfo {author} {\bibfnamefont {D.}~\bibnamefont
  {Shah}}\ and\ \bibinfo {author} {\bibfnamefont {T.}~\bibnamefont {Zaman}},\
  }\bibfield  {title} {\enquote {\bibinfo {title} {Rumors in a network: Who's
  the culprit?}}\ }\href@noop {} {\bibfield  {journal} {\bibinfo  {journal}
  {{IEEE} Trans Inf Theory}\ }\textbf {\bibinfo {volume} {57}},\ \bibinfo
  {pages} {5163{\textendash}5181} (\bibinfo {year} {2011})}\BibitemShut
  {NoStop}%
\bibitem [{\citenamefont {Fioriti}\ and\ \citenamefont
  {Chinnici}(2012)}]{fioriti2012predicting}%
  \BibitemOpen
  \bibfield  {author} {\bibinfo {author} {\bibfnamefont {V.}~\bibnamefont
  {Fioriti}}\ and\ \bibinfo {author} {\bibfnamefont {M.}~\bibnamefont
  {Chinnici}},\ }\bibfield  {title} {\enquote {\bibinfo {title} {Predicting the
  sources of an outbreak with a spectral technique},}\ }\href@noop {}
  {\bibfield  {journal} {\bibinfo  {journal} {arXiv preprint arXiv:1211.2333}\
  } (\bibinfo {year} {2012})}\BibitemShut {NoStop}%
\bibitem [{\citenamefont {Pinto}\ \emph {et~al.}(2012)\citenamefont {Pinto},
  \citenamefont {Thiran},\ and\ \citenamefont
  {Vetterli}}]{pinto_locating_2012}%
  \BibitemOpen
  \bibfield  {author} {\bibinfo {author} {\bibfnamefont {P.~C.}\ \bibnamefont
  {Pinto}}, \bibinfo {author} {\bibfnamefont {P.}~\bibnamefont {Thiran}}, \
  and\ \bibinfo {author} {\bibfnamefont {M.}~\bibnamefont {Vetterli}},\
  }\bibfield  {title} {\enquote {\bibinfo {title} {Locating the source of
  diffusion in large-scale networks},}\ }\href {\doibase
  10.1103/PhysRevLett.109.068702} {\bibfield  {journal} {\bibinfo  {journal}
  {Phys. Rev. Lett.}\ }\textbf {\bibinfo {volume} {109}},\ \bibinfo {pages}
  {068702} (\bibinfo {year} {2012})}\BibitemShut {NoStop}%
\bibitem [{\citenamefont {Antulov-Fantulin}\ \emph {et~al.}(2013)\citenamefont
  {Antulov-Fantulin}, \citenamefont {Lancic}, \citenamefont {Stefancic},
  \citenamefont {Sikic},\ and\ \citenamefont
  {Smuc}}]{antulov-fantulin_statistical_2013}%
  \BibitemOpen
  \bibfield  {author} {\bibinfo {author} {\bibfnamefont {N.}~\bibnamefont
  {Antulov-Fantulin}}, \bibinfo {author} {\bibfnamefont {A.}~\bibnamefont
  {Lancic}}, \bibinfo {author} {\bibfnamefont {H.}~\bibnamefont {Stefancic}},
  \bibinfo {author} {\bibfnamefont {M.}~\bibnamefont {Sikic}}, \ and\ \bibinfo
  {author} {\bibfnamefont {T.}~\bibnamefont {Smuc}},\ }\href
  {http://arxiv.org/abs/1304.0018} {\emph {\bibinfo {title} {Statistical
  inference framework for source detection of contagion processes on arbitrary
  network structures}}},\ \bibinfo {type} {{arXiv} e-print}\ \bibinfo {number}
  {1304.0018}\ (\bibinfo {year} {2013})\BibitemShut {NoStop}%
\bibitem [{\citenamefont {Dong}\ \emph {et~al.}(2013)\citenamefont {Dong},
  \citenamefont {Zhang},\ and\ \citenamefont {Tan}}]{dong2013rooting}%
  \BibitemOpen
  \bibfield  {author} {\bibinfo {author} {\bibfnamefont {W.}~\bibnamefont
  {Dong}}, \bibinfo {author} {\bibfnamefont {W.}~\bibnamefont {Zhang}}, \ and\
  \bibinfo {author} {\bibfnamefont {C.~W.}\ \bibnamefont {Tan}},\ }\bibfield
  {title} {\enquote {\bibinfo {title} {Rooting out the rumor culprit from
  suspects},}\ }in\ \href {\doibase 10.1109/ISIT.2013.6620711} {\emph {\bibinfo
  {booktitle} {Information Theory Proceedings (ISIT), 2013 IEEE International
  Symposium on}}}\ (\bibinfo {year} {2013})\ pp.\ \bibinfo {pages}
  {2671--2675}\BibitemShut {NoStop}%
\bibitem [{\citenamefont {Lokhov}\ \emph {et~al.}(2013)\citenamefont {Lokhov},
  \citenamefont {M{\'e}zard}, \citenamefont {Ohta},\ and\ \citenamefont
  {Zdeborov{\'a}}}]{lokhov_inferring_2013}%
  \BibitemOpen
  \bibfield  {author} {\bibinfo {author} {\bibfnamefont {A.~Y.}\ \bibnamefont
  {Lokhov}}, \bibinfo {author} {\bibfnamefont {M.}~\bibnamefont {M{\'e}zard}},
  \bibinfo {author} {\bibfnamefont {H.}~\bibnamefont {Ohta}}, \ and\ \bibinfo
  {author} {\bibfnamefont {L.}~\bibnamefont {Zdeborov{\'a}}},\ }\href
  {http://arxiv.org/abs/1303.5315} {\emph {\bibinfo {title} {Inferring the
  origin of an epidemy with dynamic message-passing algorithm}}},\ \bibinfo
  {type} {{arXiv} e-print}\ \bibinfo {number} {1303.5315}\ (\bibinfo {year}
  {2013})\BibitemShut {NoStop}%
\bibitem [{\citenamefont {Luo}\ \emph {et~al.}(2013)\citenamefont {Luo},
  \citenamefont {Tay},\ and\ \citenamefont {Leng}}]{luo_identifying_2013}%
  \BibitemOpen
  \bibfield  {author} {\bibinfo {author} {\bibfnamefont {W.}~\bibnamefont
  {Luo}}, \bibinfo {author} {\bibfnamefont {W.~P.}\ \bibnamefont {Tay}}, \ and\
  \bibinfo {author} {\bibfnamefont {M.}~\bibnamefont {Leng}},\ }\bibfield
  {title} {\enquote {\bibinfo {title} {Identifying infection sources and
  regions in large networks},}\ }\href {\doibase 10.1109/TSP.2013.2256902}
  {\bibfield  {journal} {\bibinfo  {journal} {{IEEE} Trans. Signal Process.}\
  }\textbf {\bibinfo {volume} {61}},\ \bibinfo {pages} {2850} (\bibinfo {year}
  {2013})}\BibitemShut {NoStop}%
\bibitem [{\citenamefont {Zhu}\ and\ \citenamefont
  {Ying}(2013)}]{zhu_information_2013}%
  \BibitemOpen
  \bibfield  {author} {\bibinfo {author} {\bibfnamefont {K.}~\bibnamefont
  {Zhu}}\ and\ \bibinfo {author} {\bibfnamefont {L.}~\bibnamefont {Ying}},\
  }\bibfield  {title} {\enquote {\bibinfo {title} {Information source detection
  in the {SIR} model: A sample path based approach},}\ }in\ \href {\doibase
  10.1109/ITA.2013.6502991} {\emph {\bibinfo {booktitle} {Information Theory
  and Applications Workshop ({ITA)}, 2013}}}\ (\bibinfo {year} {2013})\ pp.\
  \bibinfo {pages} {1--9}\BibitemShut {NoStop}%
\bibitem [{\citenamefont {Karamchandani}\ and\ \citenamefont
  {Franceschetti}(2013)}]{franceschetti2013}%
  \BibitemOpen
  \bibfield  {author} {\bibinfo {author} {\bibfnamefont {N.}~\bibnamefont
  {Karamchandani}}\ and\ \bibinfo {author} {\bibfnamefont {M.}~\bibnamefont
  {Franceschetti}},\ }\bibfield  {title} {\enquote {\bibinfo {title} {Rumor
  source detection under probabilistic sampling},}\ }in\ \href {\doibase
  10.1109/ISIT.2013.6620613} {\emph {\bibinfo {booktitle} {Information Theory
  Proceedings (ISIT), 2013 IEEE International Symposium on}}}\ (\bibinfo {year}
  {2013})\ pp.\ \bibinfo {pages} {2184--2188}\BibitemShut {NoStop}%
\bibitem [{\citenamefont {Altarelli}\ \emph {et~al.}(2014)\citenamefont
  {Altarelli}, \citenamefont {Braunstein}, \citenamefont {Dall'Asta},
  \citenamefont {Lage-Castellanos},\ and\ \citenamefont
  {Zecchina}}]{altarelli_prl_2014}%
  \BibitemOpen
  \bibfield  {author} {\bibinfo {author} {\bibfnamefont {F.}~\bibnamefont
  {Altarelli}}, \bibinfo {author} {\bibfnamefont {A.}~\bibnamefont
  {Braunstein}}, \bibinfo {author} {\bibfnamefont {L.}~\bibnamefont
  {Dall'Asta}}, \bibinfo {author} {\bibfnamefont {A.}~\bibnamefont
  {Lage-Castellanos}}, \ and\ \bibinfo {author} {\bibfnamefont
  {R.}~\bibnamefont {Zecchina}},\ }\bibfield  {title} {\enquote {\bibinfo
  {title} {Bayesian inference of epidemics on networks via belief
  propagation},}\ }\href {\doibase 10.1103/PhysRevLett.112.118701} {\bibfield
  {journal} {\bibinfo  {journal} {Phys. Rev. Lett.}\ }\textbf {\bibinfo
  {volume} {112}},\ \bibinfo {pages} {118701} (\bibinfo {year}
  {2014})}\BibitemShut {NoStop}%
\bibitem [{\citenamefont {Karrer}\ and\ \citenamefont
  {Newman}(2010)}]{karrer_message_2010}%
  \BibitemOpen
  \bibfield  {author} {\bibinfo {author} {\bibfnamefont {B.}~\bibnamefont
  {Karrer}}\ and\ \bibinfo {author} {\bibfnamefont {M.~E.~J.}\ \bibnamefont
  {Newman}},\ }\bibfield  {title} {\enquote {\bibinfo {title} {Message passing
  approach for general epidemic models},}\ }\href {\doibase
  10.1103/PhysRevE.82.016101} {\bibfield  {journal} {\bibinfo  {journal} {Phys.
  Rev. E}\ }\textbf {\bibinfo {volume} {82}},\ \bibinfo {pages} {016101}
  (\bibinfo {year} {2010})}\BibitemShut {NoStop}%
\bibitem [{\citenamefont {Milling}\ \emph {et~al.}(2013)\citenamefont
  {Milling}, \citenamefont {Caramanis}, \citenamefont {Mannor},\ and\
  \citenamefont {Shakkottai}}]{milling2013detecting}%
  \BibitemOpen
  \bibfield  {author} {\bibinfo {author} {\bibfnamefont {C.}~\bibnamefont
  {Milling}}, \bibinfo {author} {\bibfnamefont {C.}~\bibnamefont {Caramanis}},
  \bibinfo {author} {\bibfnamefont {S.}~\bibnamefont {Mannor}}, \ and\ \bibinfo
  {author} {\bibfnamefont {S.}~\bibnamefont {Shakkottai}},\ }\bibfield  {title}
  {\enquote {\bibinfo {title} {Detecting epidemics using highly noisy data},}\
  }in\ \href@noop {} {\emph {\bibinfo {booktitle} {Proceedings of the
  fourteenth ACM international symposium on Mobile ad hoc networking and
  computing}}}\ (\bibinfo {organization} {ACM},\ \bibinfo {year} {2013})\ pp.\
  \bibinfo {pages} {177--186}\BibitemShut {NoStop}%
\bibitem [{\citenamefont {Meirom}\ \emph {et~al.}(2014)\citenamefont {Meirom},
  \citenamefont {Milling}, \citenamefont {Caramanis}, \citenamefont {Mannor},
  \citenamefont {Orda},\ and\ \citenamefont
  {Shakkottai}}]{meirom2014localized}%
  \BibitemOpen
  \bibfield  {author} {\bibinfo {author} {\bibfnamefont {E.~A.}\ \bibnamefont
  {Meirom}}, \bibinfo {author} {\bibfnamefont {C.}~\bibnamefont {Milling}},
  \bibinfo {author} {\bibfnamefont {C.}~\bibnamefont {Caramanis}}, \bibinfo
  {author} {\bibfnamefont {S.}~\bibnamefont {Mannor}}, \bibinfo {author}
  {\bibfnamefont {A.}~\bibnamefont {Orda}}, \ and\ \bibinfo {author}
  {\bibfnamefont {S.}~\bibnamefont {Shakkottai}},\ }\bibfield  {title}
  {\enquote {\bibinfo {title} {Localized epidemic detection in networks with
  overwhelming noise},}\ }\href@noop {} {\bibfield  {journal} {\bibinfo
  {journal} {arXiv preprint arXiv:1402.1263}\ } (\bibinfo {year}
  {2014})}\BibitemShut {NoStop}%
\bibitem [{\citenamefont {Kermack}\ and\ \citenamefont
  {McKendrick}(1932)}]{kermack1932contributions}%
  \BibitemOpen
  \bibfield  {author} {\bibinfo {author} {\bibfnamefont {W.~O.}\ \bibnamefont
  {Kermack}}\ and\ \bibinfo {author} {\bibfnamefont {A.~G.}\ \bibnamefont
  {McKendrick}},\ }\bibfield  {title} {\enquote {\bibinfo {title}
  {Contributions to the mathematical theory of epidemics. ii. the problem of
  endemicity},}\ }\href@noop {} {\bibfield  {journal} {\bibinfo  {journal}
  {Proceedings of the Royal society of London. Series A}\ }\textbf {\bibinfo
  {volume} {138}},\ \bibinfo {pages} {55} (\bibinfo {year} {1932})}\BibitemShut
  {NoStop}%
\bibitem [{\citenamefont {Altarelli}\ \emph
  {et~al.}(2013{\natexlab{a}})\citenamefont {Altarelli}, \citenamefont
  {Braunstein}, \citenamefont {{Dall'Asta}},\ and\ \citenamefont
  {Zecchina}}]{altarelli_large_2013}%
  \BibitemOpen
  \bibfield  {author} {\bibinfo {author} {\bibfnamefont {F.}~\bibnamefont
  {Altarelli}}, \bibinfo {author} {\bibfnamefont {A.}~\bibnamefont
  {Braunstein}}, \bibinfo {author} {\bibfnamefont {L.}~\bibnamefont
  {{Dall'Asta}}}, \ and\ \bibinfo {author} {\bibfnamefont {R.}~\bibnamefont
  {Zecchina}},\ }\bibfield  {title} {\enquote {\bibinfo {title} {Large
  deviations of cascade processes on graphs},}\ }\href {\doibase
  10.1103/PhysRevE.87.062115} {\bibfield  {journal} {\bibinfo  {journal} {Phys.
  Rev. E}\ }\textbf {\bibinfo {volume} {87}},\ \bibinfo {pages} {062115}
  (\bibinfo {year} {2013}{\natexlab{a}})}\BibitemShut {NoStop}%
\bibitem [{\citenamefont {Altarelli}\ \emph
  {et~al.}(2013{\natexlab{b}})\citenamefont {Altarelli}, \citenamefont
  {Braunstein}, \citenamefont {{Dall{\textquoteright}Asta}},\ and\
  \citenamefont {Zecchina}}]{altarelli_optimizing_2013}%
  \BibitemOpen
  \bibfield  {author} {\bibinfo {author} {\bibfnamefont {F.}~\bibnamefont
  {Altarelli}}, \bibinfo {author} {\bibfnamefont {A.}~\bibnamefont
  {Braunstein}}, \bibinfo {author} {\bibfnamefont {L.}~\bibnamefont
  {{Dall{\textquoteright}Asta}}}, \ and\ \bibinfo {author} {\bibfnamefont
  {R.}~\bibnamefont {Zecchina}},\ }\bibfield  {title} {\enquote {\bibinfo
  {title} {Optimizing spread dynamics on graphs by message passing},}\ }\href
  {\doibase 10.1088/1742-5468/2013/09/P09011} {\bibfield  {journal} {\bibinfo
  {journal} {J. Stat. Mech.}\ }\textbf {\bibinfo {volume} {2013}},\ \bibinfo
  {pages} {P09011} (\bibinfo {year} {2013}{\natexlab{b}})}\BibitemShut
  {NoStop}%
\bibitem [{\citenamefont {Barab\'asi}\ and\ \citenamefont
  {Albert}(1999)}]{Barabasi15101999}%
  \BibitemOpen
  \bibfield  {author} {\bibinfo {author} {\bibfnamefont {A.-L.}\ \bibnamefont
  {Barab\'asi}}\ and\ \bibinfo {author} {\bibfnamefont {R.}~\bibnamefont
  {Albert}},\ }\bibfield  {title} {\enquote {\bibinfo {title} {Emergence of
  scaling in random networks},}\ }\href {\doibase 10.1126/science.286.5439.509}
  {\bibfield  {journal} {\bibinfo  {journal} {Science}\ }\textbf {\bibinfo
  {volume} {286}},\ \bibinfo {pages} {509} (\bibinfo {year} {1999})},\ \Eprint
  {http://arxiv.org/abs/http://www.sciencemag.org/content/286/5439/509.full.pdf}
  {http://www.sciencemag.org/content/286/5439/509.full.pdf} \BibitemShut
  {NoStop}%
\bibitem [{\citenamefont {Yedidia}\ \emph {et~al.}(2001)\citenamefont
  {Yedidia}, \citenamefont {Freeman},\ and\ \citenamefont
  {Weiss}}]{yedidia2001bethe}%
  \BibitemOpen
  \bibfield  {author} {\bibinfo {author} {\bibfnamefont {J.~S.}\ \bibnamefont
  {Yedidia}}, \bibinfo {author} {\bibfnamefont {W.~T.}\ \bibnamefont
  {Freeman}}, \ and\ \bibinfo {author} {\bibfnamefont {Y.}~\bibnamefont
  {Weiss}},\ }\bibfield  {title} {\enquote {\bibinfo {title} {Bethe free
  energy, kikuchi approximations, and belief propagation algorithms},}\
  }\href@noop {} {\bibfield  {journal} {\bibinfo  {journal} {Advances in neural
  information processing systems}\ }\textbf {\bibinfo {volume} {13}} (\bibinfo
  {year} {2001})}\BibitemShut {NoStop}%
\end{thebibliography}

%
\end{document}